\documentclass[]{aa} 
\usepackage{graphicx}
\usepackage{amsmath} 
\usepackage{amssymb} 
\usepackage{placeins}
\usepackage{txfonts}
\usepackage{ulem}
\usepackage{comment}
\usepackage{caption}
\usepackage{subcaption}

\newcommand{\be}{\begin{equation}}
\newcommand{\ee}{\end{equation}}
\newcommand{\ba}{\begin{eqnarray}}
\newcommand{\ea}{\end{eqnarray}}
\newcommand{\lla}{\left\langle}
\newcommand{\rra}{\right\rangle}
\newcommand{\vect}[1]{{\boldsymbol{#1}}}
\newcommand{\ddr}[1]{\frac{\mathrm{d}}{\mathrm{d}R}}
\newcommand{\dr}[1]{\frac{\mathrm{d}{#1}}{\mathrm{d}R}}
\newcommand{\dre}{\frac{\mathrm{d}}{\mathrm{d}R}}
\newcommand{\drl}[1]{\mathrm{d}{#1}/\mathrm{d}R}
\newcommand{\drle}{\mathrm{d}_R}

\newcommand{\grad}{ {\bf \nabla } }
\newcommand{\curl}{ {\bf \nabla} \times}
\newcommand{\divv}{ {\bf \nabla} \cdot}

\begin{document}

\title{Shear interaction and acceleration of a corotating stream detected by Parker Solar Probe.}

\author{
Andrea Verdini\inst{1,2}
\corrauth{andrea.verdini@unifi.it}
\and
Simone Landi\inst{1,2}
\email{simone.landi@unifi.it}
\and
Emanuele Papini\inst{3,4}
\email{emanuele.papini@inaf.it}
\and
Luca Franci\inst{4,3}
\email{luca.franci@northumbria.ac.uk}
\and
Petr Hellinger\inst{5,6}
\email{petr.hellinger@asu.cas.cz}
\and
Roberto Livi\inst{7}
\email{rlivi@ssl.berkeley.edu}
\and
Orlando Romeo\inst{7}
\email{oromeo@ssl.berkeley.edu}
}

\institute{
Dipartimento di Fisica e Astronomia, Universit\'a degli Studi di Firenze, Firenze, Italy
\and
INAF Osservatorio Astrofisico di Arcetri, Firenze, Italy
\and
Istituto di Astrofisica e Planetologia Spaziali (IAPS), Istituto Nazionale di Astrofisica (INAF), Roma, Italy
\and
School of Engineering, Physics and Mathematics, Northumbria University, Newcastle upon Tyne, UK
\and
Institute of Atmospheric Physics of the Czech Academy of Sciences, Prague, Czech Republic
\and
Astronomical Institute of the Czech Academy of Sciences, Prague, Czech Republic
\and
Space Sciences Laboratory, University of California, Berkeley, CA, USA
}

\date{Accepted XXX. Received YYY; in original form ZZZ}

\abstract{
Recent analysis of Parker Solar Probe (PSP) and Solar Orbiter data indicates that fluctuations energy is transferred to the bulk flow in fast streams of the solar wind at heliocentric distances between 15 and 120$R_\odot$.
To gain insight into this process, we analyze data collected during the inbound orbit of PSP in Encounter 10, when it is in approximate corotation with the Sun and an accelerating stream is detected between 25 and 45$R_\odot$.
The geometry of the flow indicates that a neighboring very fast streams is colliding with the corotating stream, causing large-scale magnetic field distortion, density enhancement, and acceleration. However, also deposition of fluctuations' energy can heat and do work on the solar wind plasma, resulting in a radial acceleration. We evaluate the energy lost by fluctuations via radial variation of conserved quantities in the corotating stream and find that it is almost equally partitioned into heating and work. While the heating is marginally consistent with the non-adiabatic decrease of proton temperature, the work exerted on the wind is insufficient to account for the measured solar wind acceleration.
Although limited to this event, observations suggest that shear interaction is capable of accelerating slower streams within relatively short distances from the Sun, possibly leaving its imprint as large-scale density and magnetic fluctuations.
}


\maketitle
\nolinenumbers

\section{Introduction}
In the solar wind, magnetic and velocity fluctuations carry substantial energy that can be transferred to plasma as heat,
increasing the pressure gradients and indirectly accelerating the solar wind, or as work, i.e. directly accelerating the wind with momentum deposition.
One dimensional solar wind models allow to understand how energy must be distributed to achieve fast solar wind streams \citep[e.g.][]{Leer:1982vq,Hansteen:2012up}.
Close to the Sun, heat deposition is necessary to reach coronal temperatures around or above million K and accelerates the wind to supersonic speed.
The flux tube expansion controls the distribution of heat around the sonic point and thus the resulting mass flux \citep[e.g.][]{Wang:1993vo}.
Further out, momentum deposition in the supersonic flow allows reaching large speed of about 700km/s, characteristic of fast streams, without altering the mass flux. 

Beyond 0.3au in-situ data show a non-adiabatic decay of proton temperature and an acceleration of slow speed streams \citep{Totten:1995ve,Maksimovic:2020aa}, indicating the need of heat and momentum deposition also at that distances.
Transfer of fluctuations' energy to small scales by a turbulent cascade is compatible with the required heating, as indicated by data analysis and numerical simulations \citep{Marino:2008uu,Stawarz:2009tq,Montagud-Camps:2018we,Montagud-Camps:2020vw}, but the specific heating mechanism is yet unknown.
Observations closer to the Sun became available only recently with the advent of Parker Solar Probe (PSP, \citealt{Fox:2016aa}) and Solar Orbiter (SolO, \citealt{Muller:2020aa}).
A statistical analysis on the first 10 PSP orbits confirmed theoretical expectations: for distances between 15 and 90 $R_\odot$ the acceleration of fast streams is achieved at the expenses of fluctuations energy, while for slower streams the driver is the ambipolar electric field \citep{Halekas:2023uc}.
\citet{Bourouaine:2024aa} analysed a statistical sample of Alfv\'enic slow streams detected by PSP and SolO between 0.1 and 1au. Measuring the damping of energy flux as a function of distance they concluded that the energy lost by fluctuations was compatible with both proton and electron heating, and also consistent with phenomenological turbulent heating for incompressible transverse fluctuations \citep{Chandran:2009tu}.
The identification of the same plasma parcel during the alignment of SolO and PSP in encounter 9, allowed \citet{Rivera:2024aa} to compare the mass and energy flux  at distances of $15$ and $120~R_\odot$ while the solar wind speed was passing from 400 to 500km/s.
Total energy flux was conserved only when accounting for the energy in fluctuations. 
Their energy loss was almost equally partitioned into work on the flow and heat of the plasma.
Finally, the estimated solar wind acceleration was compatible with the sum of wave pressure and thermal pressure gradients and 
the heating from fluctuations was compatible with that required to explain the observed non-adiabatic temperature decrease.
The authors concluded that fluctuations energy is necessary to explain the solar wind acceleration and heating in this range of distances.
This conclusion is based on a simplified one-dimensional model of the solar wind \citep{Dakeyo:2022aa,Shi:2022ab}, which offers a necessary but simple theoretical background to interpret data.
One may ask whether the identified plasma parcel could have travelled for such a large distance in isolation and without significant interaction with the surrounding solar wind.
In this respect, intervals in which PSP is in corotation with the Sun allows analyzing radial variation of the solar wind on smaller extents, possibly giving the opportunity to sample isolated streams that originate from the same stationary source.

\citet{Badman:2023aa} identified three distinct coronal holes as the sources of three different streams detected by PSP during Encounter 10.
During its inbound orbit PSP was in quasi-corotation with the Sun and measured plasma emanating from a narrow region on the Sun surface (labelled stream 1). 
The magnetic field extrapolation combined with back-mapping of streams detected by PSP indicated that stream 1 was not uniform: it was mostly composed of a collection of fast streams emanating from the equatorial extension of the
southern polar coronal hole \citep{Shi:2020aa}, with the remainder slower stream connected to the longitudinal extension of the coronal hole boundary (or possibly to a neighboring coronal hole).
This composite structure was somehow overlooked in subsequent papers that interpreted PSP data corresponding to stream 1 as the radial sampling of a single stream.
\citet{Davis:2023th} analyzed the spectra of magnetic fluctuations in 12hr-long intervals which showed a substantial evolution with distance.
At scales smaller than about a minute the spectrum had a typical power-law index close to Kolmogorov, at every distance.
At larger scales instead the power-law index of the frequency spectrum passed from -0.5 to -1 when distance varied from $\approx15$ to $\approx45R_\odot$, a behavior confirmed by statistical analysis \citep{Huang:2023aa}.
Such large scales were analyzed in \citet{Bowen:2025aa} who showed that they consist of switchbacks, i.e. circularly polarized large amplitude fluctuations, whose  amplitude and tilt relative to the mean magnetic field were increasing with distance. 
Their energy instead decreased in a way consistent with damping of the wave action density, that is, deposition of heat.
Third-order moments returned a cascade rate comparable to both the heating obtained from the decay of double-adiabatic invariants and from the above-mentioned turbulent phenomenology,  confirming the statistical findings \citep{Bourouaine:2024aa} that identified turbulent cascade as the process responsible for the plasma heating.

In this work we focus again on stream 1 recorded during the inbound orbit of Encounter 10, while PSP was in quasi-corotation with the Sun, to estimate the momentum deposition by fluctuations and understand if it is compatible with the solar wind acceleration.
Data and their processing are described in section~\ref{sec:data}.
 We re-examine the stream structure in section~\ref{sec:stream} and find that it is unlikely that PSP is detecting the radial evolution of an isolated stream. 
Consistently with \citet{Badman:2023aa}, we identify a very fast stream and a slower stream (that we label corotating stream) and suggest that the acceleration of the latter is caused by their shear interaction.
In section~\ref{sec:heat}, in a way similar to \citet{Bourouaine:2024aa}, we compute the work exerted by fluctuations on the corotating stream and the decrease of their energy flux with distance to obtain the heating that result from the loss of fluctuations energy.
We conclude that shear interaction, and not deposition of fluctuations energy, is probably responsible for the acceleration of the corotating stream.
In section~\ref{sec:discuss} we test our conclusion against changes in the length of the interval and the timescale of the analysis and 
in the final section we summarize our results and discuss their limitations and implications.

\section{Data}\label{sec:data}
We analyze magnetic field and plasma data collected by PSP during Encounter 10, from 16-11-2021 00:00 UTC to 21-11-2021 00:00 UTC.
We use Level 2 magnetic field data in RTN coordinates from Fluxgate Magnetometer \citep{Bale:2016aa}, electron density and temperature obtained from Quasi Thermal Noise spectroscopy \citep{Moncuquet:2020aa} in the FIELDS experiment, level 3 proton plasma moments in RTN coordinates and temperature tensor from Solar Probe Analyzer for Ions (SPAN-I) \citep{Livi:2022aa}, part of the Solar Wind Electrons Alphas and Protons (SWEAP) suite \citep{Kasper:2016aa}.
For each set of instrument data, we first interpolate measurements on a regular grid, replacing only non consecutive NaN and bad data with their linearly interpolated value.
Magnetic field measurements are rebinned on proton resolution with a boxcar smoothing with window equal to the proton time increment.

\begin{figure}[t!]
\includegraphics[width=0.98\linewidth,trim={1.9cm 5.5cm 1.8cm 1.6cm},clip=]{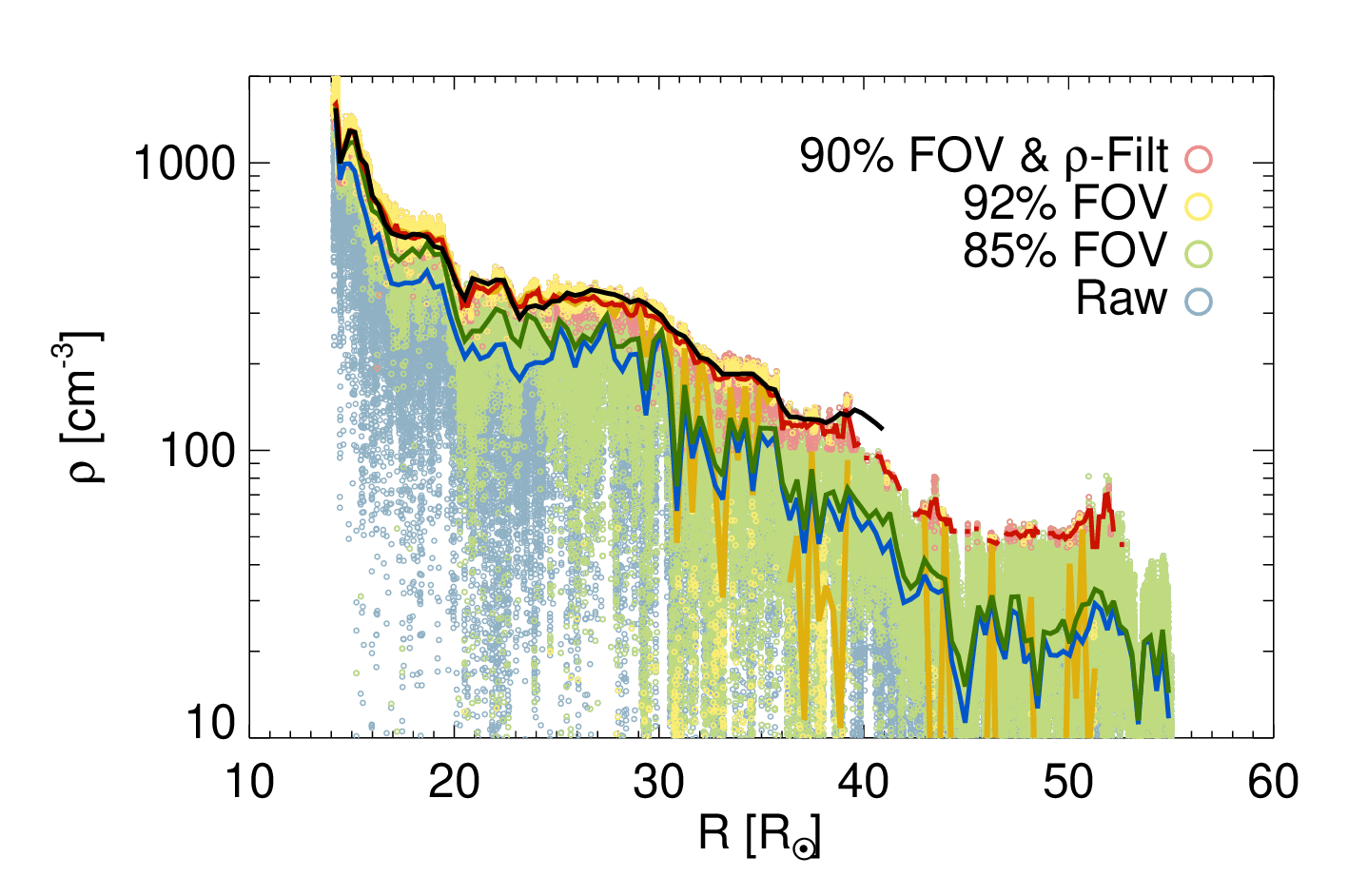}\\
\includegraphics[width=0.98\linewidth,trim={1.9cm 1.3cm 1.8cm 2.cm},clip=]{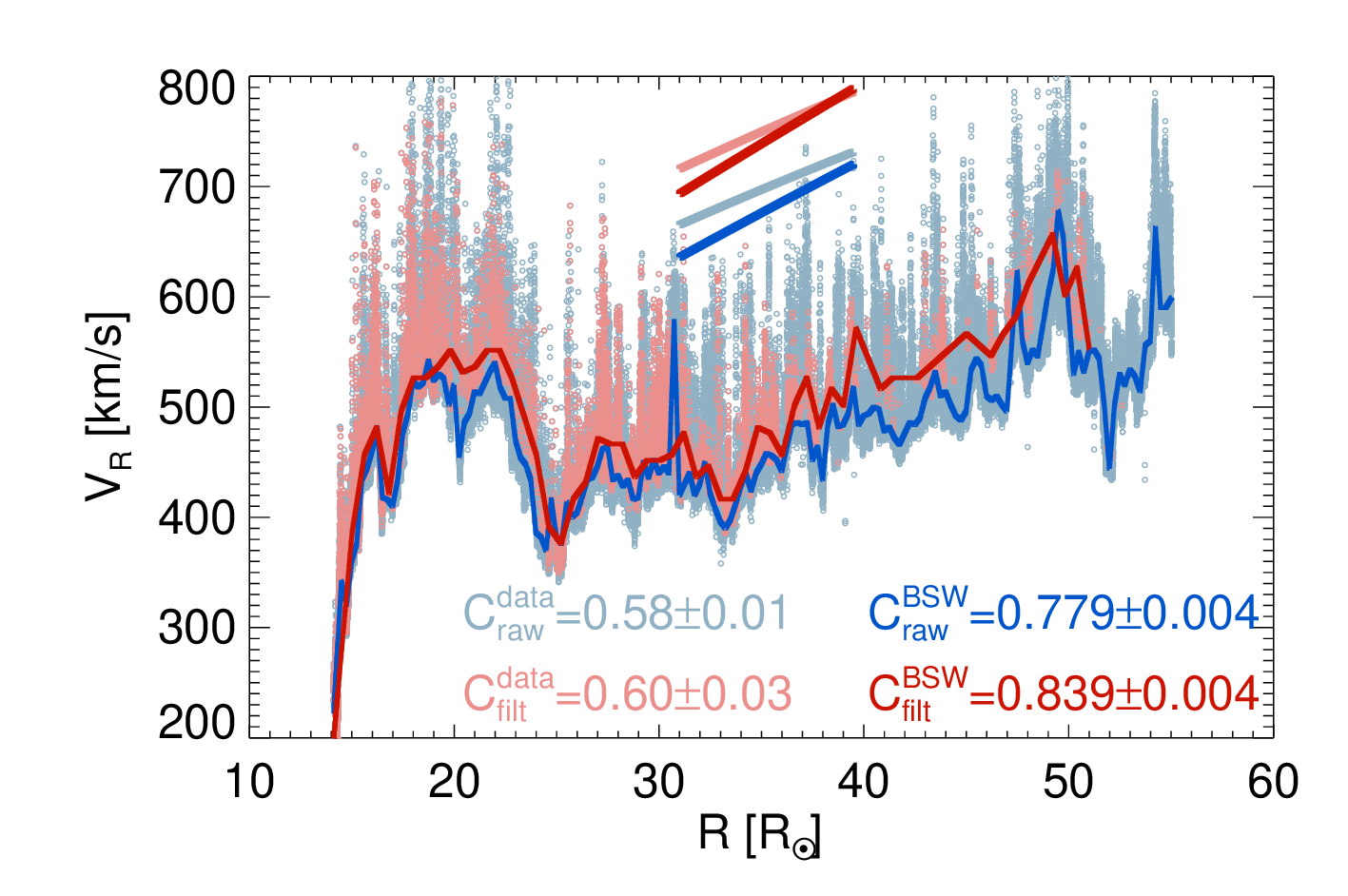}

\caption{Top panel. Proton density ($\rho$) as a function of distance from SPAN-I moments obtained with raw data (light blue empty circles), or filtering out records that have a  coverage of the instrument FOV smaller than 85\% and 92\% (green and yellow empty circles, respectively). Red empty circles are the records retained in our analysis. Thick lines with darker colors are the corresponding averages over non-overlapping 1~hr intervals. The average electron density from QTN is also plotted for reference with a black line.
Bottom panel. Records of radial velocity $V_R$ in the raw and filtered dataset (blue and red empty circles, respectively) and the corresponding mode value (background solar wind - BSW). 
Thicker lines in the upper part are power law fit to data (raw and filtered) and to the BSW (raw and filtered) in the range $31,~40~R_\odot$.
The corresponding power-law indexes are reported on the figure.}
\label{fig:modal}
\end{figure}

Since moments form SPAN-I instrument may suffer from incomplete collection of the velocity distribution function in the field of view (FOV) of the instrument, 
we fit a simplified gaussian distribution to the data and evaluate the FOV coverage with the portion of the fitted function that lies within the FOV range of the instrument \citep{Romeo:2024aa}.
The top panel of Fig.~\ref{fig:modal} shows with light blue, light green, and light yellow circles the proton density as a function of distance for the raw records, for 85\%, and for 92\% coverage, respectively. 
The corresponding proton density averaged on 1hr scale ($\approx0.5R_\odot$) is also plotted with solid lines (same color scale, but darker).
For reference, the average electron density from QTN measurements is overplotted with a black line.
As the FOV coverage is increased, records with lower and lower densities are excluded, making the filtered sample and its mean closer to the electron density. 
However, for the highest coverage, the number of records at distances larger than about 35~$R_\odot$ decreases dramatically and few records display anomalously small density, specifically smaller than 100 and 50~$\mathrm{cm^{-3}}$ below and above 40$R_\odot$, respectively, producing strong oscillations in the corresponding average density for $R\gtrsim30~R_\odot$.
As a trade-off in term of quality and quantity of records, we thus lower to 90\% the threshold in the FOV coverage and further filter out records with anomalous small density as identified in the 92\% FOV coverage. This represent our final dataset and is shown with red circles. The average proton density, plotted with dark red line, is almost identical to the average of the 92\% coverage records for $R\lesssim29R_\odot$, and is very close to the average electron density for $R\lesssim40~R_\odot$.

The bottom panel of fig.~\ref{fig:modal} compares the radial wind speed in the raw records (light blue circles) and in the filtered records (light red circles).
Data gaps appear in the latter for distances larger than $40~R_\odot$.
With solid blue and red lines we plot the mode of $V_R$ computed in bins of $0.25$ and $0.6R_\odot$ for the raw records and filtered records, respectively.
The mode is somehow representative of the ``background'' solar wind (BSW), i.e. the velocity profile that the wind would have in absence of switchbacks.
Switchbacks are clearly visible as strong one-sided radial fluctuations in both raw and filtered records and they are known to positively contribute to the average solar wind speed \citep{Matteini:2014ud,Bourouaine:2022aa,Agapitov:2023aa}.

A natural question when estimating the solar wind acceleration is wether we should consider switchbacks in evaluating the wind speed and use averages, or instead exclude them and use the BSW.
In sect.~\ref{sec:heat}, the solar wind acceleration $a_{sw}=V_R\drle V_R$ will be measured estimating the derivative via power law fit on the wind speed, $V_R\propto R^C$, for distances between $\approx31$ and $40~R_\odot$.
The power law coefficients, $C$, are displayed in the figure: subscripts $raw$ and $filt$, indicate if raw or filtered records are use, while superscripts $data$ and $BSW$ indicate a fit on the full dataset or the BSW. For clarity the fitted profiles are plotted with a different stack constant in the raw and filtered cases.
When using data, the coefficients for raw and filtered case are very similar and consistent (light blue and light red symbols). 
When fitting the BSW, instead, the filtered case yields a larger coefficient (dark blue and dark red lines), indicating that records with smaller $V_R$ are progressively removed by the filtering in this range of distances.
For both raw and filtered records, the fit on the data returns a smaller coefficient than that of the corresponding BSW. Since the fit on data include the contribution of switchbacks, their amplitude must decrease as a function of distance, suggesting a transfer of energy to the bulk flow. 
However a larger coefficient does not necessarily imply a larger solar wind acceleration.
For this reason, we will use both the BSW and the average solar wind to estimate upper and lower limits of $a_{sw}$ in sec~\ref{sec:heat}.

In the following, using the rebinned and filtered data, we compute the mean and root-mean-squared values (rms) of each field, typically at a timescale $\tau=1~\mathrm{hr}$, 
shifting the averaging window by 20 minutes to increase statistics. 
As a rule, capital letters are used for the mean fields, small letters for fluctuations.
Power-law fit will be used to estimate the radial derivative of mean quantities, fluctuations energy flux, and some terms entering the work of fluctuations.
We will use one-sigma uncertainty for the fitted coefficients and represent the absolute error (not the geometric mean of errors) of dependent variables as shaded areas.
In sect.~\ref{sec:discuss} both the range of distances and the timescale of averages will be varied to further estimate uncertainties on the results.

\begin{figure*}[t!]
\includegraphics[width=\columnwidth,trim={0.5cm 0.2cm 0.4cm 0.6cm},clip=]{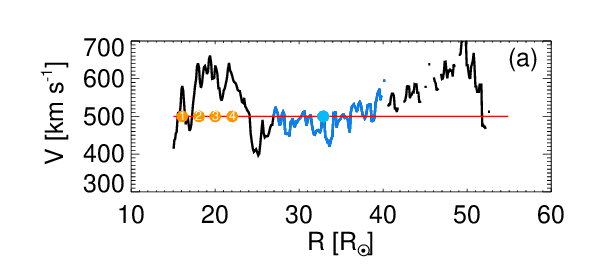}
\includegraphics[width=\columnwidth,trim={0.5cm  0.2cm 0.4cm 0.6cm},clip=]{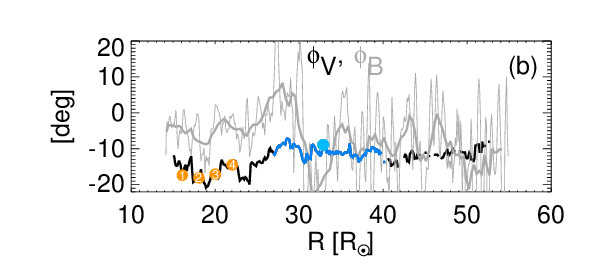}\\
\includegraphics[width=\columnwidth,trim={0.5cm 0cm 0.35cm 1.8cm},clip=]{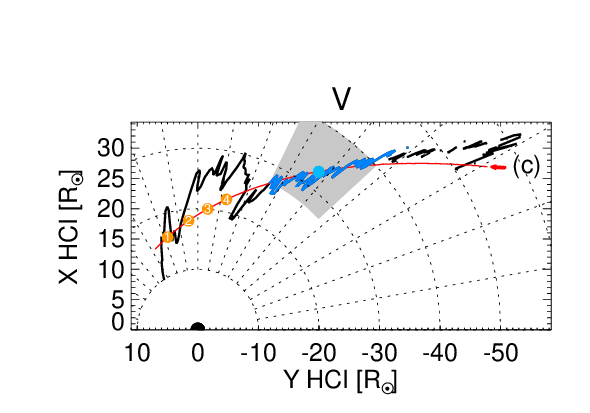} 
\includegraphics[width=\columnwidth,trim={0.5cm 0cm 0.35cm 1.8cm},clip=]{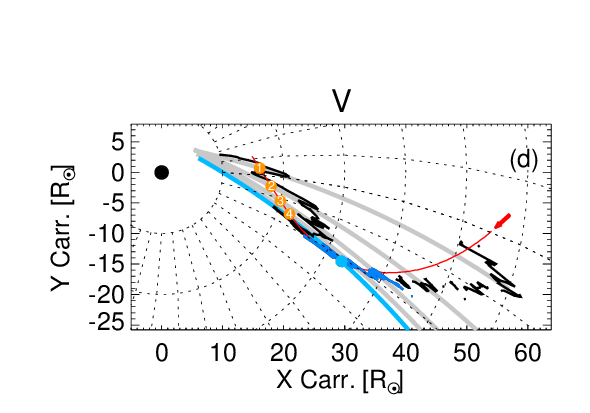} 
\caption{Overview of the the stream structure during the inbound orbit of Encouter 10. In each panel the period of corotation is highlighted with a thick blue line and  dots mark five selected PSP locations, $R_s$ (the blue dot corresponding to strict corotation).
(a) Radial variation of the stream velocity ($V$) and reference velocity $V_{ref}$ (red line).
(b) Azimuthal tilt angle of the solar wind direction ($\phi_V$, black) and of the magnetic field direction ($\phi_B$, grey). The latter is also computed with a 6hr moving window with 1hr shift and plotted with a thick grey line.
(c) Parker Solar Probe position, $\vect{R_{PSP}}$ (red), and solar wind position, $\vect{R_{SW}}$ (black), projected on the solar equatorial plane in the fixed Heliocentric Inertial system (HCI). The grey shaded area indicate the region of corotation. Note that x and y axis are flipped.
(d) Same as in panel (c) but in the frame rotating with the Sun (Carrington coordinates). Streamlines for a wind at constant $V_{ref}$ flowing radially at $10R_\odot$ are plotted with dotted lines. Gray and bleu lines trace streamlines of constant speed passing from $R_s$ with the measured velocity $V(R_s)$ and tilt angle $\phi_V(R_S)$.}
\label{fig:orbit}
\end{figure*}

\section{Structure and properties of the stream}\label{sec:stream}
The stream configuration is summarized in fig.~\ref{fig:orbit}, where we use quantities that are averaged on 1~hr. 
In each panel the blue lines highlight quasi corotation, defined as $V_{T,PSP}/\omega_\odot R_{PSP} \in[2/3,3/2]$, where $R_{PSP}$ and $V_{T,PSP}$ are the position and tangential speed of PSP, and $\omega_\odot$ is the Sun's angular speed. 
For reference, strict corotation is marked with a blue dot at $R\approx33~R_\odot$ and 4 streams are labelled with numbered orange dots.
In panel (a), the solar wind speed is plotted as a function of distance. 
It shows a non-monotonic profile, with a very fast stream appearing in the first $25R_\odot$, followed by a slower stream that is accelerating from 450 to 700~$\mathrm{km~s^{-1}}$ at a distance of about 50~$R_\odot$. The data gap between 40 and 45~$R_\odot$ is due to the poor FOV coverage in the SPAN-I instrument, the horizontal red line is a reference speed, $V_{ref}=500~\mathrm{km~s^{-1}}$, that help visualizing the polar plots of panels (c) and (d).
As suggested by \citet{Horbury:2023aa}, the large-scale modulation of stream velocity on scales of about $3R_\odot$ seems to be associated with switchback patches on the same scale that are clearly seen in magnetic and velocity in timeseries (not shown here, see for example \citealt{Shi:2020aa}).
The azimuthal tilt angle with respect to the radial direction is shown in panel (b) as a function of distance both for the flow direction $\phi_V$ (black) and for the magnetic field direction, $\phi_B$ (grey).
Large oscillations are seen in the magnetic field vector because of the large relative magnetic fluctuations, $\delta B/B_0\approx1$. 
For easier comparison with the velocity direction, we also compute $\phi_B$ by averaging over $\tau=6~\mathrm{hr}$ on a moving window with 1hr shift (thick grey line).
In the whole interval the flow and magnetic field vectors are tilted in the direction opposite to the solar rotation.
Below $25R_\odot$ the two vectors are misaligned by about $15^o$, with the flow vector being oblique ($-20^o$) and the magnetic filed almost radial ($-5^o$). 
For $R\in[25,30]R_\odot$ the two vectors turn in the direction of the Sun's rotation, while keeping their offset, and at larger distances they become basically aligned with each other with an orientation of $-10^o$.

The geometry of the encounter is shown in panel (c) in a fixed reference frame centered on the Sun, the Heliocentric Inertial System - HCI (note that x and y axis are flipped), and in panel (d) in a frame rotating with the Sun, employing Carrington coordinates. 
The PSP position, $\vect{R_{PSP}}$,  is plotted with a red line. Superimposed with a black line is the solar wind ``position'', $\vect{R_{SW}}=\vect{R_{PSP}}+\Delta R\vect{\hat{V}_{SW}}$, which, loosely speaking, is the projection on the PSP orbit of the velocity profile in panel (a) accounting for the flow direction in panel (b).
More precisely, the vector $\vect{R_{SW}}$ is a kinematic propagation in the direction of the stream, $\vect{\hat{V}_{SW}}$, of the plasma detected at the PSP orbit, with the travelled distance proportional to the relative variation of the wind speed, $\Delta R=R_{ref}(V/V_{ref}-1)$, where $R_{ref}\approx36R_\odot$ is arbitrarily chosen to magnify the flow structure. 

In the fixed HCI frame (panel c), the inclination of streams is barely visible in the corotation zone (shaded area) and further out, but the fast stream detected in the closest approach is clearly tilted in direction opposite to solar rotation. 
Corotation spans about $15R_\odot$ in distance and less than $30^o$ in longitude, lasting for about 30~hr. 

In panel (d) the encounter is shown in the reference frame rotating with the Sun.
To trace streamlines we use a simple kinematic model in which energy and angular momentum are conserved: 
a point mass launched with velocity $v_0$ and tilt angle $\phi_{v0}$ from a spinning plate of radius $R_0$.
Dashed lines are reference streamlines for a uniform source: they are drawn considering flows separated by $10^o$ with radial velocity $V_{ref}$ ($\phi_{v0}=0$) at $R_0=10~R_\odot$.
They suggest that fast streams recorded at the beginning and at the end of PSP orbit originate from a source occupying $\gtrsim20^o$ in longitude and during propagation they do not mix with each other.
A slightly more accurate representation of streamlines is obtained when the launch velocity and tilt angle ($v_0$ and $\phi_{v0}$) are traced back from PSP measurements.
We select four locations corresponding to fast streams with large tilt angles, $R_s=16,~18,~20,~22~R_\odot$ (numbered orange filled circles), and also the position corresponding to corotation, $R_s=33~R_\odot$ (blue filled circle).
According to the grey streamlines, streams 1 to 4 originate from a common localized source ($\lesssim10^o$ wide at 6~$R_\odot$) and the corotating stream flows from from the longitudinal extent of the same source (blue line). 
This is qualitatively in agreement with the mapping and source-sourface extrapolation back to the Sun \citep{Shi:2020aa,Badman:2023aa}.
While it is probable that stream 1 is connected to the fastest stream at $50~R_\odot$ and propagates away from the corotating stream, 
the proximity of the stream 3, 4 with the corotating stream and the behavior of the the magnetic field angle suggest a possible ongoing shear interaction among them.

\begin{figure}[ht]
\includegraphics[width=\columnwidth,trim={0.4cm  1.3cm 0.4cm 0.3cm},clip=]{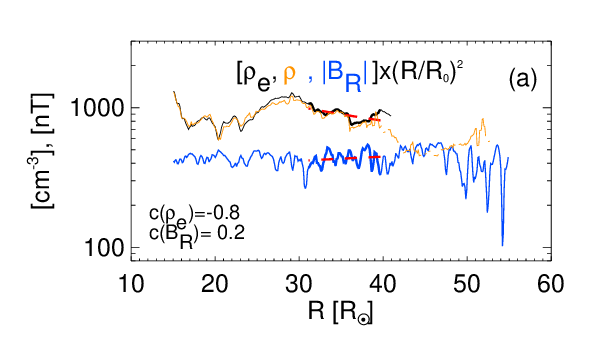}\\
\includegraphics[width=\columnwidth,trim={0.5cm 1.3cm 0.4cm 0.6cm},clip=]{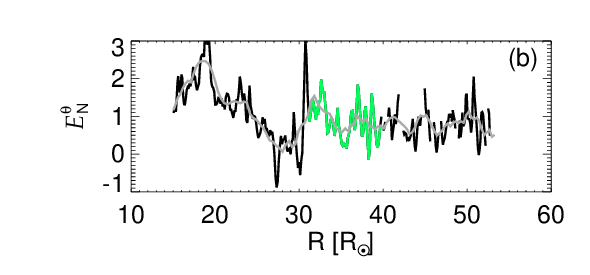}\\
\includegraphics[width=\columnwidth,trim={0.5cm 1.3cm 0.4cm 0.6cm},clip=]{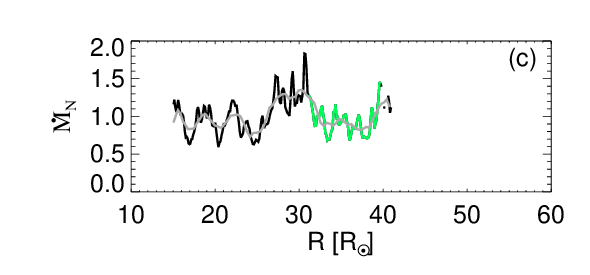}\\
\includegraphics[width=\columnwidth,trim={0.5cm 0.3cm 0.4cm 0.6cm},clip=]{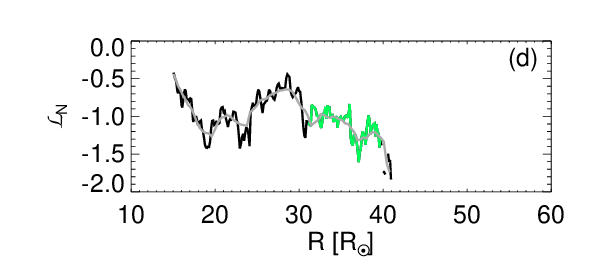} 
\caption{
Panel (a). Electron density ($\rho_e$, black), proton density ($\rho$, orange) and absolute value of the radial magnetic field ($|B_R|$, blue), compensated by $R^2$. The corotation period is indicated with thicker lines.
Three conserved quantities, Eqs.~\eqref{eq:Maxwell}-\eqref{eq:Lmom}, are plotted as a function of distance $R$ after normalization by the absolute value of their mean in the whole distance range: the normalized polar electric field, $E^\theta_N$,  in panel (b); the normalize mass flux, $\dot{M}_N$, in panel (c); the normalized angular momentum, ${\cal L}_N$, in panel (d). The interval selected for the analysis is indicated in green. The grey line is computed with $\tau=6~\mathrm{hr}$ and shifting the window by 1~hr to increase statistics.
}
\label{fig:cons}
\end{figure}

We recall that for a uniform source, streamlines have instead an azimuthal symmetry. Following \citet{Weber:1967aa}, requiring stationary density, azimuthal magnetic filed, and azimuthal velocity one obtains three conserved quantities, namely, the polar electric field, the mass flux, and the angular momentum, respectively defined as:
\ba
E^\theta&=&\sqrt{A}(V_RB_T-V_TB_R)=(V_RB_T-V_TB_R)/\sqrt{B_R},\label{eq:Maxwell}\\
\dot{M}&=&\rho V_R A=\rho V_R/B_R,\label{eq:Mflux}\\
{\cal L}&=&\sqrt{A}(V_T-\frac{B_R}{4\pi\rho V_R}B_T)=(V_T-\frac{B_R}{4\pi\rho V_R}B_T)/\sqrt{B_R}, \label{eq:Lmom}
\ea
where in the last equalities the area of the the flux-tube cross section follows from the conservation of the magnetic flux, $B_RA=const$
(in Appendix~\ref{sec:app1} we show that the three conserved quantities are unchanged when fluctuations are Alfv\'enic or perpendicular).
We inspect these quantities to understand if there is a range of distances in which the corotating stream is isolated, that is, it is propagating undisturbed without shear interaction and the observed velocity increase is a genuine radial acceleration.
We thus focus on distances larger than $25~R_\odot$.
In the top panel of Fig.~\ref{fig:cons} we show, compensated by $R^2$, the radial trend of electron density, proton density, and radial magnetic field intensity with black, orange, and blue lines, respectively.
The electron density from QTN is very close to the proton density from SPAN-I for $R\lesssim40R_\odot$. For $R\in[20,30]R_\odot$ they both display a relative density increase followed by a rapid decrease $\approx R^{-2.8}$.
Magnetic field data are more regular in the whole interval and consistent with a sub-spherical expansion, $B_R\sim R^{-1.8}$.
The conserved quantities, normalized by the modulus of their average value in the whole domain, $E^\theta_N=E^\theta/\lla E^\theta\rra,~\dot{M}_N=\dot{M}/\lla\dot{M}\rra$ and $\mathcal{L}_N=\mathcal{L}/\lla \mathcal{L}\rra$ are plotted in panels (b), (c), and (d), respectively.
The polar electric field is positive and approximately constant (it has the largest relative variation), with an excursion to negative values in correspondence of the density bump. Its profile is consistent with that of the magnetic field tilt angle, given that $B_R<0$ and $\partial_tB_\phi=-1/\sqrt{A}\drle E_\theta$.
The mass flux is better conserved, with relative variation $\lesssim 20\%$, but it clearly increases around the density bump.
Finally, the angular momentum is negative, because the flow is tilted in direction opposite to the Sun's rotation, and similarly to the mass flux it is approximately conserved between 31 and $40~R_\odot$.
Ultimately, because of the large relative variations due to the presence of switchbacks patches and streams on scales of about 4-6~hr, we cannot clearly identify any large range of distances in which all three quantities are conserved.
However, relying on the density profile, on the smoothed conserved quantities plotted with a gray line\footnote{We use as before a  running average on a 6h-window with 1h-shift.}, and on the tilt angles in fig~\ref{fig:orbit}(b), we can distinguish two regions.

In the range $R\in[25,~30]R_\odot$ the fast stream is colliding with the neighboring slower stream causing plasma compression and magnetic field bending. These are responsible for an increase of the mass flux and decrease of the polar electric field.
The density enhancement and magnetic field distortion, the concurrent acceleration of the corotating stream and deceleration of the neighboring fast stream, are all signatures of shear interaction \citep{Grappin:1996wd,Landi:2006wo,Shi:2020aa} and suggest it as the main driver of the observed stream configuration.

\begin{figure*}[ht]
 \includegraphics[width=0.49\linewidth,trim={0.4cm 1.4cm 0.4cm 0.2cm},clip=]{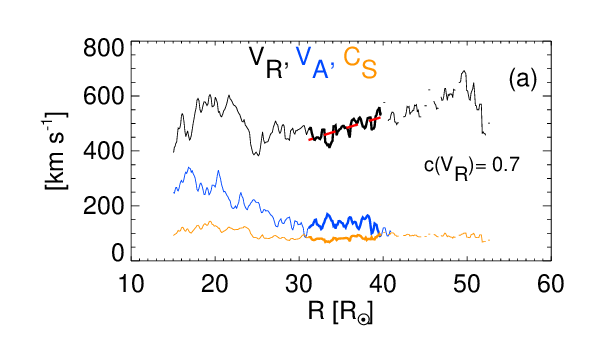}
 \includegraphics[width=0.49\linewidth,trim={0.4cm 1.4cm 0.4cm 0.2cm},clip=]{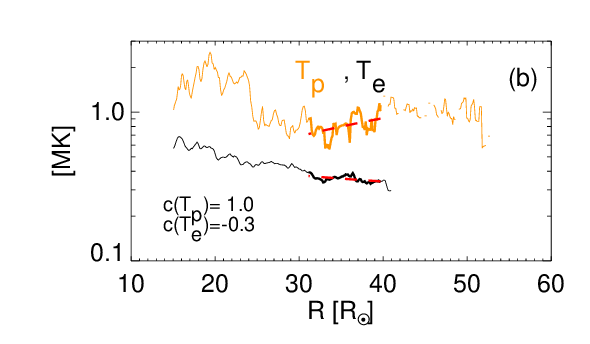}\\
 \includegraphics[width=0.49\linewidth,trim={0.5cm  0.cm 0.4cm 0.2cm},clip=]{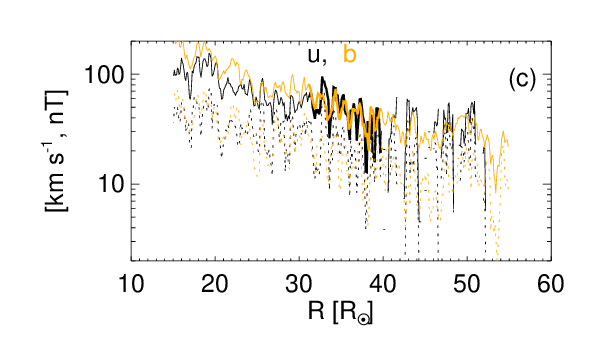}
 \includegraphics[width=0.49\linewidth,trim={0.5cm 0.cm 0.4cm 0.2cm},clip=]{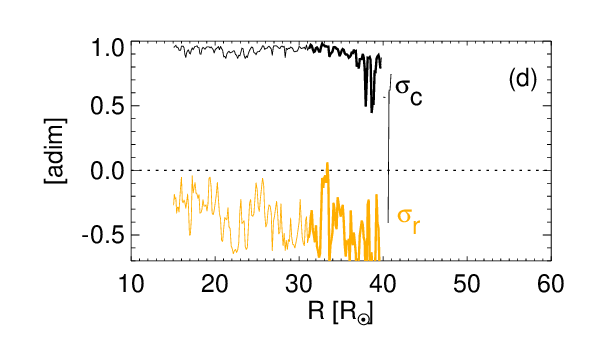}
 \caption{Properties of the solar wind stream as a function of radial distance obtained by averaging over 1h intervals. 
Quasi corotation is indicated with thick lines. Power law fit are plotted with dashed red lines and the corresponding exponents are indicated in the figure.
From left to right and top to bottom the panels show:
(a) radial solar wind speed ($V_R$, black), Alfv\'en speed ($V_A$, blue) and proton sound speed ($C_S$, orange);
(b) proton temperature ($T_p$, orange) and electron temperature ($T_e$ black);
(c) amplitudes of velocity fluctuations (black) and magnetic fluctuations (orange), the amplitude of their radial component is also plotted with dotted lines;
(d) normalized cross helicity (black) and normalized residual energy (orange).
}
\label{fig:meanflow1}
\end{figure*}

However, in the range $R\in[31,~40]R_\odot$, the approximate conservation of polar electric field, mass flux, and angular momentum suggests that this corotating stream (highlighted in green) may propagate undisturbed.
In the next section we will compute losses of fluctuations energy in the corotating stream to understand if its acceleration can be explained without invoking shear interaction. 
Before that, we conclude the analysis of the radial variation of the flow and fluctuations properties in Fig.~\ref{fig:meanflow1},
where the corotation stream is highlighted with thick lines.
In panel (a) the radial wind speed, Alfv\'en velocity, and proton sound speed are plotted in black, blue, and orange line, respectively.
The radial flow is largely super-Alfv\'enic and supersonic everywhere, the plasma beta is around 0.1 at short distances and approaches unity at about $40R_\odot$. Beyond 40~$R_\odot$ the absence of Alfv\'en speed data is due to missing (or very few) QTN density measurements, while the paucity of points in the velocity and proton temperature reflects the large amount of discarded data according to FOV coverage.
The corotating stream is accelerating, the velocity increase is estimated via a power law fit yielding a coefficient 0.7 as indicated in figure.
Combining it with the power-law indexes of the radial magnetic field and electron density reported in Fig.~\ref{fig:cons}a we get $\dot{M}\propto R^{-0.3}$, on average the mass flux is slightly decreasing.
In panel (b) the proton temperature (orange line) follows approximately the velocity profile, with larger temperature corresponding to larger speed and showing again large-scale modulations. Note that the corotating stream is heated, with an estimate power-law index 1. 
On the contrary, the electron temperature (black line) decreases smoothly in the whole interval, with approximately a unique power-law.
Radial trends of fluctuations amplitudes are shown in panel (c) with solid lines, dotted lines refer to the radial components (also very similar to the field-parallel components).
Velocity fluctuations (black lines) are marginally subsonic everywhere, their amplitude decreases up to about $30R_\odot$ and then increases again because of the contribution of the radial component. 
On the contrary, the amplitude of magnetic fluctuations (orange lines) and of their radial component decreases almost monotonically (except for $R\gtrsim45$).
In panel (d) we plot the normalized cross-helicity,
 $\sigma_c=-2\vect{u}\cdot\vect{b}/\sqrt{4\pi\rho}/(u^2+b^2/4\pi\rho)$ and the normalized residual energy,
 $\sigma_r=(u^2-b^2/4\pi\rho)/(u^2+b^2/4\pi\rho)$, with black and orange lines, respectively.
Below 30~$R_\odot$ the cross helicity is large and about constant with a value around 0.9, then it drops to 0.8 in the corotating stream because of the decoupling between $b_R$ and $u_R$. 
The residual energy displays large and small-scale oscillations but is negative and decreasing: since density is strongly decreasing the magnetic energy becomes more and more important with distance, except at 34~$R_\odot$ where a small scale density enhancment occurs.
 
To summarize, if the source can be considered stationary during the inbound phase of PSP orbit (about 2 days), kinematically modeled streamlines suggest that the outer part of the corotating stream is connected to fast streams that originate from the same extended source, located at about 6~$R_\odot$ and with a size comparable to the scale of supergranulation ($10^o$). 
According to this interpretation, the radial acceleration of the corotating stream is due to shear interaction. 
However, in the same range of distances the quasi conservation of mass flux, angular momentum, and polar electric field suggests the presence of a uniform stream that is genuinely accelerating at the expenses of fluctuations energy.
In the next section we use mean and rms amplitude of fluctuations to quantify their energy loss as a function of distance and to estimate the resulting proton heating and stream acceleration.

\section{Heating and acceleration in the corotating stream}\label{sec:heat}
To obtain the contribution of fluctuations to heating and acceleration of the mean flow, we assume mean fields are stationary and depend only on the radial coordinate, $R$.
Fluctuations are statistically stationary and incompressible, density and (proton and electron) pressure have vanishing fluctuations and are uniquely identified by their mean values, $\rho$ and $P$.
For simplicity we report only results relevant to the data analysis, while a complete derivation is given in Appendix~\ref{sec:app1} and follows closely the work of \citet{Hollweg:1974aa}.
For radially aligned flow and mean magnetic field directions, the energy equation for fluctuations reads,
\be
Q_{f}=-\left[W_f+\frac{1}{A}\dr{(AF_f)}\right]\label{eq:qtot}
\ee
where $Q_{f}>0$ is the volumetric heating that results from damping ($-Q_f$) of the fluctuations energy, $W_f$ is the work done by fluctuations on the mean flow, and $F_f$ is the radial flux of fluctuations energy.
We consider here two simplified models that allow us to compute $W_f$ and $F_f$, and thus the heating, directly from rms and mean values (their full expressions can be found in eqs.~\eqref{eqa:flux}-\eqref{eqa:work} of appendix~\ref{sec:app1}). 
The Alfv\'enic model (subscript $Alf$) is valid for strictly correlated (or anitcorrelated) fluctuations, $\vect{u}=\pm\vect{b}/\sqrt{4\pi\rho}$, that maintain a locally constant magnetic field, $\delta|\vect{B}|=0$, without any restriction on components or wave-vectors (the original derivation in \citealt{Hollweg:1974aa}):
\ba
W_{Alf} &=& -U_{R}\frac{1}{8\pi}\dr{\lla b^2\rra}, \label{eq:a_alf}\\
F_{Alf}&=&U_{R}\frac{3}{2}\frac{\lla b^2\rra}{4\pi}+
    |V_{aR}|\frac{\lla b^2\rra}{4\pi} \label{eq:f_alf}.
\ea
The index $R$ stands for the radial component of vectors, $V_{aR}=B_R/\sqrt{4\pi\rho}$ is the mean Alfv\'en speed, and its absolute value is used to yield the correct sign for unidirectional outward fluctuations (resulting in positive or negative v-b correlation when  $B_{R}<0$ or $>0$, respectively).
The perpendicular model (subscript $\bot$) has fluctuations perpendicular to the mean flow and magnetic field ($\vect{u}=\vect{u_\bot}$, $\vect{b}=\vect{b_\bot}$) but decorrelation is allowed (i.e. they represent a superposition of counter-propagating fluctuations):
\ba
W_\bot &=& -U_{R}\frac{1}{8\pi}\dr{\lla b_\bot^2\rra}
+U_R\frac{A'}{2A} \left(\rho\lla u_\bot^2\rra -\frac{\lla b_\bot^2\rra}{4\pi}\right),\label{eq:a_bot}\\
F_{\bot}&=&U_{R}\left[\frac{\rho\lla u_\bot^2\rra}{2}+\frac{\lla b_\bot^2\rra}{4\pi}\right]
   -B_{R}\frac{\vect{u_\bot}\cdot\vect{b_\bot}}{4\pi}.
\label{eq:f_bot}
\ea
Now $B_{R}$ is kept with its sign with respect to $R$, for dominant anti-sunward propagating fluctuations the last term in the flux is always positive.

\begin{figure}[t!]
 \includegraphics[width=\linewidth,trim={0.0cm  0.cm 0.cm 0cm},clip=]{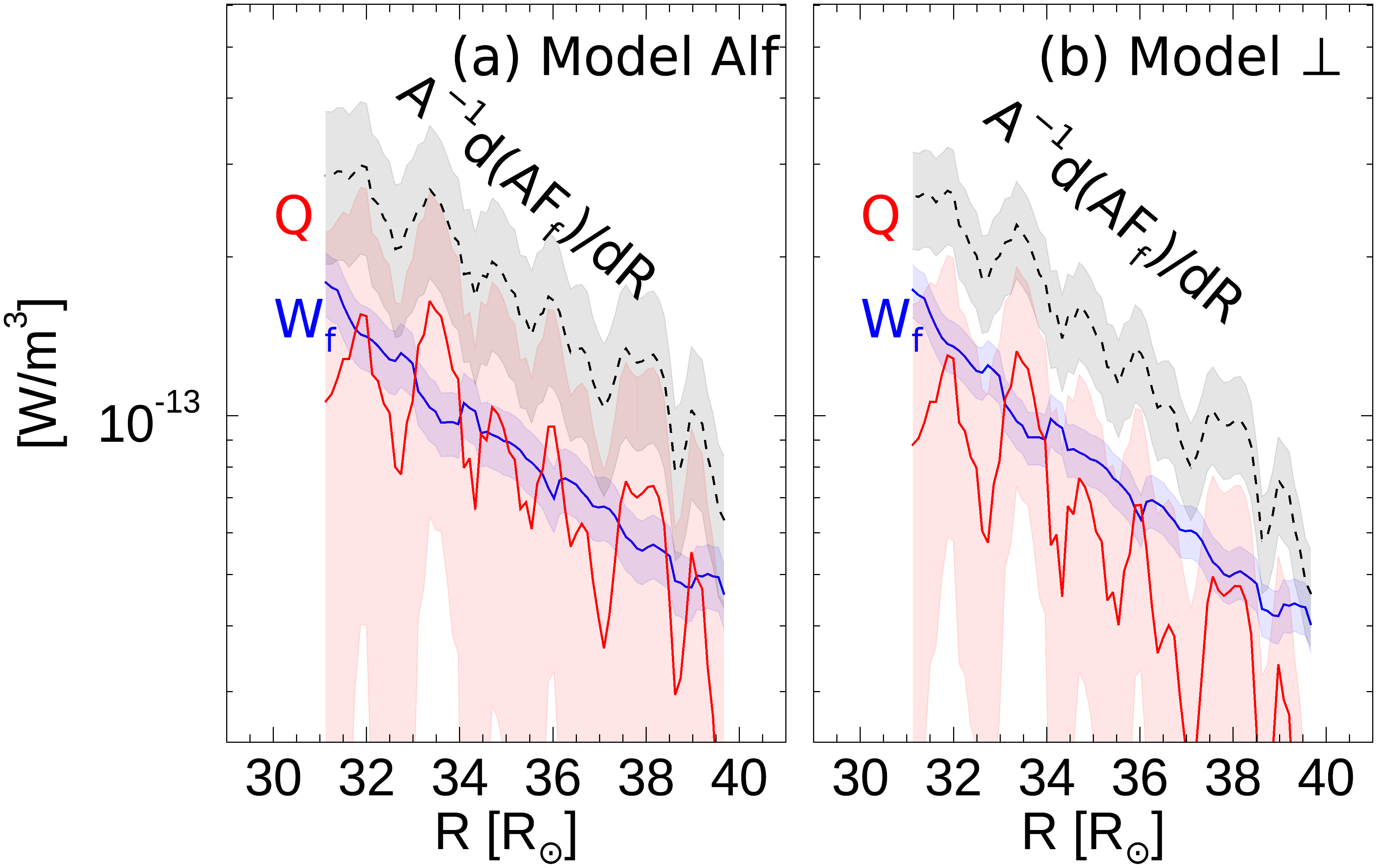}
\caption{Variation of the flux of fluctuations energy ($A^{-1}\drl{(AF_f)}$, black line), work on the mean flow ($W_f$, blue line), and heating ($Q_f$, red line) as a function of distance.
Solid (dashed) lines represent positive (negative) values, shaded areas represent uncertainties. 
In panel (a) we use the Alfv\'enic model for $W_f$ and $F_f$, eqs.~\eqref{eq:a_alf}-\eqref{eq:f_alf}. In panel (b) we use the perpendicular model,  eqs.~\eqref{eq:a_bot}-\eqref{eq:f_bot}.}
\label{fig:loss}
\end{figure}

Eqs.~\eqref{eq:qtot}, along with the two model equations, eqs.~\eqref{eq:a_alf}-\eqref{eq:f_bot}, is our estimate of the volumetric heating $Q_{f}>0$ from conservation equations. 
In these expressions we use the QTN electron density, $\rho=\rho_e$, assumed to be equal to the proton density. 
The flux tube cross section obeys magnetic flux conservations, $AB_{R}=const$ and we compute derivatives via a power-law fit, with uncertainties given by the $1\sigma$ deviation of the fitted coefficients. 
In panel (a) and (b) of Fig.~\ref{fig:loss} we compare the Alfv\'enic and perpendicular model by plotting the derivative of the energy flux, the work, and the resulting heating as a function of distance.
Dashed and solid lines are used for negative and positive quantities, respectively, with shaded areas representing the uncertainty.
Both models return very similar profiles of the work (blue line), with the Alfv\'enic model yielding a larger variation of the energy flux derivative and consequently a larger volumetric heating.
On average, about 50\% of energy losses are spent to heat the plasma in the Alfv\'enic model, and 40\% in the perpendicular model.

These heating rates obtained from conservation equations are now compared to the expected proton heating, $Q_p$, and to a phenomenological turbulent heating, $Q_{DC}$ \citep{Dmitruk:2002us,Chandran:2009tu},
\be
Q_{DC}=-\rho\frac{1+V_{aR}/U_{R}}{V_{aR}/U_{R}}\frac{z_+^2}{4}\dr{V_{aR}},
\label{eq:q_dc}
\ee
where $\vect{z_\pm}=\vect{u_\bot}\mp\vect{b_\bot}/\sqrt{4\pi\rho}$ are the Elsasser fields obtained with only perpendicular components
 \footnote{This phenomenology was first introduced by \citet{Dmitruk:2002us} for a static atmosphere, and later extended by \citet{Chandran:2009tu} to the expanding solar wind. It is obtained for strongly imbalanced incompressible transverse fluctuations, say $z_+\gg z_-$, when the amplitude of $z_-$ is set by the equilibrium between nonlinear turbulent damping and linear reflection terms, i.e. in the low frequency limit.}.
The expected proton heating $Q_p$, is evaluated in two ways.
From the isotropic temperature equation neglecting heat conduction, we have
\be
Q_{iso}=k_BU_{R}\left[\frac{3}{2}n\dr{T_{p}}-T_{p}\dr{n}\right],\label{eq:qpe}
\ee
where we used $\gamma=5/3$ and $k_B$ is the Boltzman constant.
In the double adiabatic approximation, we have $Q_{ad}=Q_\|+Q_\bot$, where 
\ba
Q_\|&=&- \lla P_\|\rra U_{R}\frac{\mathrm{d}\log\lla P_\|B^2/n^3\rra}{\mathrm{d}R},\label{eq:qpar}\\
Q_\bot&=&-\lla P_\bot\rra U_{R}\frac{\mathrm{d}\log\lla P_\bot/nB\rra}{\mathrm{d}R},\label{eq:qperp}
\ea
are the parallel and perpendicular heating, evaluated from the radial variations of adiabatic invariants
\citep[e.g.][]{Hellinger:2013tw,Zaslavsky:2023aa,Bowen:2025aa}.
We compute parallel and perpendicular pressures, $P_\|$, $P_\bot$, by first rotating the temperature tensor along the 1hr mean field $\vect{B_0}$ and then by averaging. $B$ is the averaged magnetic field amplitude and $n$ is the mean proton density from SPAN-I data.
\begin{figure}[t]
\includegraphics[width=\linewidth,trim={0.cm 0.cm 0.cm 0cm},clip=]{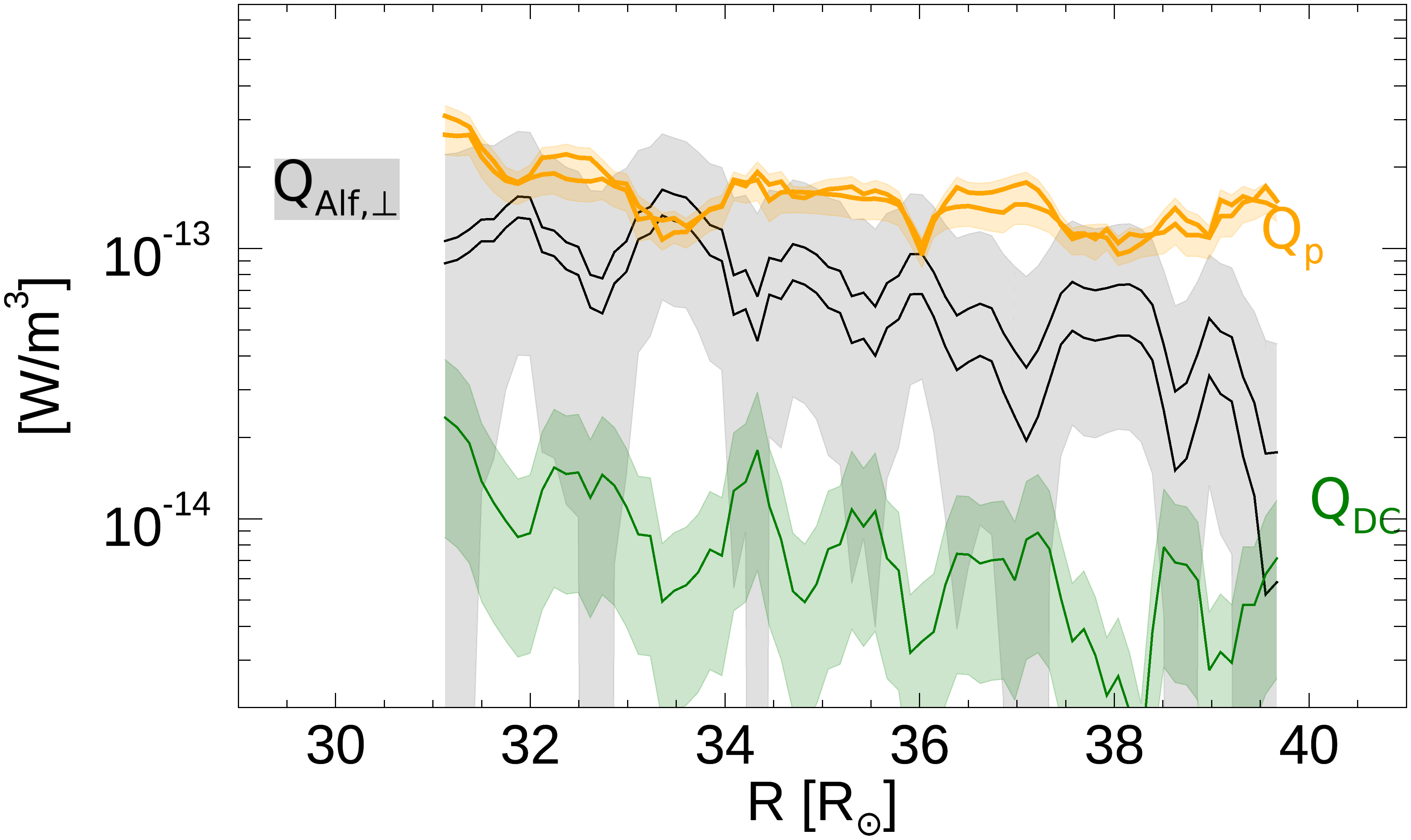}
\caption{Volumetric heating as obtained from conservation equations in the Alfv\'en and $\bot$ models, $Q_{Alf,\bot}$ (black lines), and from Dmitruk-Chandran turbulence closure, $Q_{DC}$ (geen line). 
The expected proton heating, $Q_p$ (orange), is bounded by the two estimates in the isotropic and double adiabatic approximations, eqs.~\eqref{eq:qpe} -\eqref{eq:qperp}.
Shaded areas represent  uncertainties.}
\label{fig:heat}
\end{figure}

In Fig.~\ref{fig:heat}, we plot as a function of distance the heating from conservation equations, $Q_{f}=Q_{Alf,\bot}$ (black lines), the phenomenological turbulent heating, $Q_{DC}$ (green line), and the expected proton heating, $Q_p$ (orange lines).
Shaded areas represent uncertainties.
Both $Q_{Alf}$ and $Q_\bot$ along with their uncertainties are used to determine upper and lower bounds of $Q_{f}$.
Analogously, $Q_{iso}$ and $Q_{ad}$ determine upper and lower bounds of the expected proton heating $Q_p$.
The heating estimated with conservation equations is barely sufficient to maintain the observed temperature profile, $Q_{Alf,\bot}\approx Q_p$, becoming slightly smaller at large distances, while the phenomenological heating is a factor 10 below the required values in the whole interval. 
This is partially in contrast with results of \citet{Bowen:2025aa}, who reported for the same encounter $Q_{DC}\approx Q_p\approx Q_Y$, with $Q_Y$ estimated from third-order moments (but see the discussion section), and the statistical analysis of \citet{Bourouaine:2024aa}.

\begin{figure}[t!]
 \includegraphics[width=\linewidth,trim={0.cm 0.cm 0.cm 0cm},clip=]{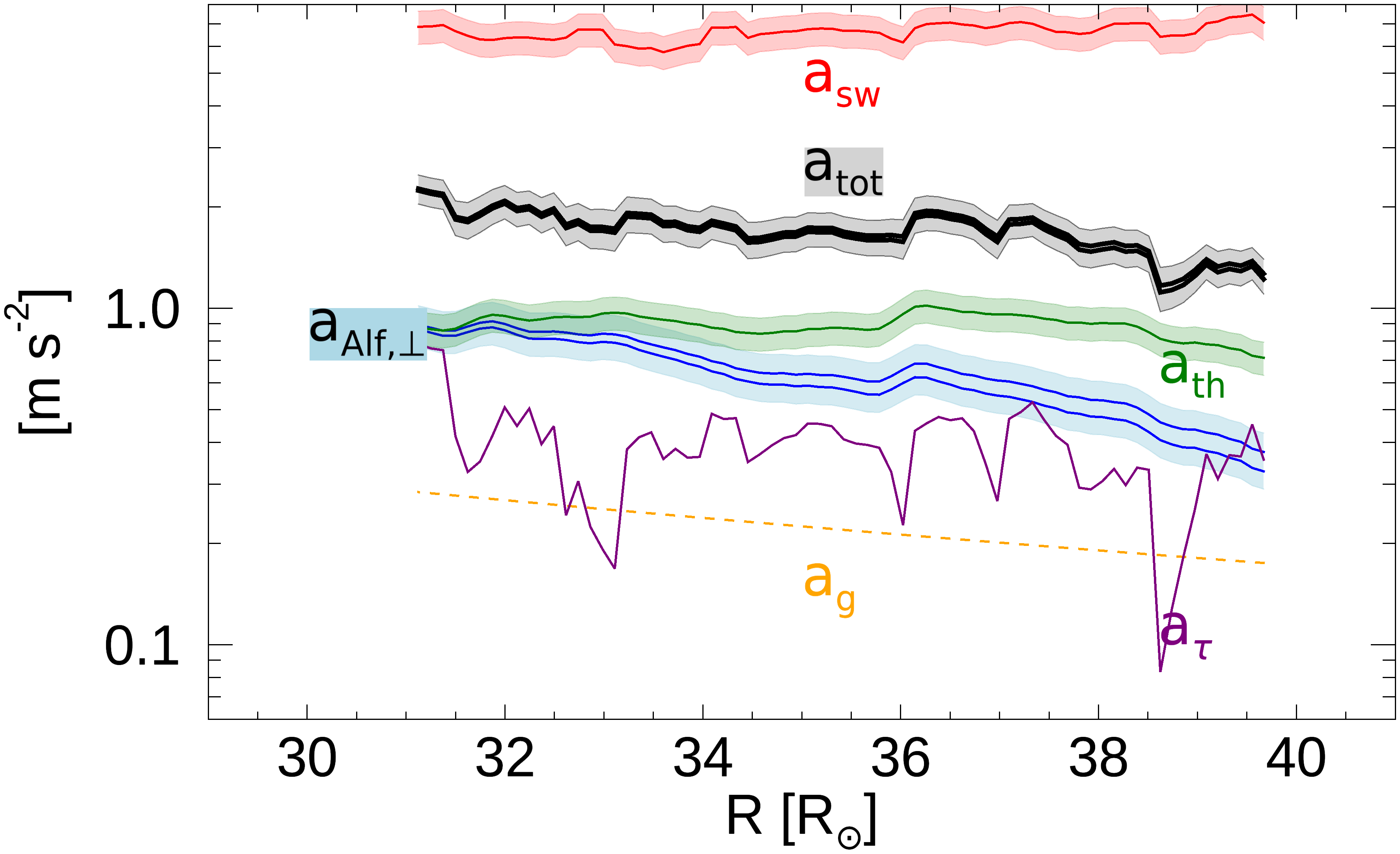}
\caption{The solar wind acceleration, $a_w$ (red line) is plotted as a function of distance and compared to the total acceleration, $a_{tot}$ (black line), given by all terms on the r.h.s of eq.~\eqref{eq:asw}: 
the thermal contribution, $a_{th}$ (green line); 
the fluctuations' contribution, $a_{Alf,\bot}$ (blue lines);
the gravitational term, $a_g$ (orange line);
the velocity and magnetic torque, $a_\tau$ (purple line).
Shaded areas indicate uncertainties,
positive (negative) values are plotted with solid (dashed) lines.}
\label{fig:acc}
\end{figure}

We finally use the stationary momentum equation to evaluate the contribution of the work done by fluctuations in accelerating the corotating stream,
\ba
\underbrace{U_{R}\dr{U_{R}}}_{a_{sw}}&=&\underbrace{-\frac{GM_\odot}{R^2}}_{a_g}\underbrace{-\frac{1}{\rho}\dr{(P_p+P_e)}}_{a_{th}} +\big[a_f\big]
\label{eq:asw}\\
&+&\underbrace{\left[\frac{A'}{2A}\left(U_{T}^2-\frac{B_{T}^2}{4\pi\rho}\right) - \frac{1}{8\pi\rho}\dr{B_{T}^2}\right]}_{a_\tau}. \nonumber
\ea 
On the l.h.s we have the solar wind acceleration ($a_{sw}$), while on the r.h.s. the first, second, and third term correspond to the deceleration caused by the Sun's gravity ($a_g$), the thermal acceleration exerted by proton and electron pressure ($a_{th}$), the (model-dependent) acceleration exerted by fluctuation on the mean flow ($a_f=W_f/\rho U_{R}$). 
The last term is the acceleration exerted by the non-radial flow and magnetic field ($a_\tau$).
We use QTN density to evaluate both electron and proton pressures, $P_{p,e}=nk_BT_{p,e}$.
The contribution to the solar wind acceleration of the above terms is shown in Fig~\ref{fig:acc} as a function of distance.
The work done by fluctuations (blue) contributes to the acceleration of the solar wind with almost the same amount of thermal pressure, which, we recall, is sustained with heating by fluctuations.
Both are larger than the acceleration caused by non radial mean flow and magnetic field (purple) and the gravitational deceleration (yellow).
However, the sum of all these terms, $a_{tot}$ (black line), is at least a factor 3 smaller that the measured acceleration of the stream, $a_{sw}$ (red line, computed on the average $V_R$ and on the BSW).
In conclusion, despite fluctuations accelerate the stream both directly (work) and indirectly (heating), the measured acceleration cannot be explained by their contribution only and, generally speaking, 2D effects must come into play.

\begin{figure*}[ht!]
\includegraphics[height=0.4\linewidth,trim={0.1cm .1cm 0cm 0.cm},clip=]{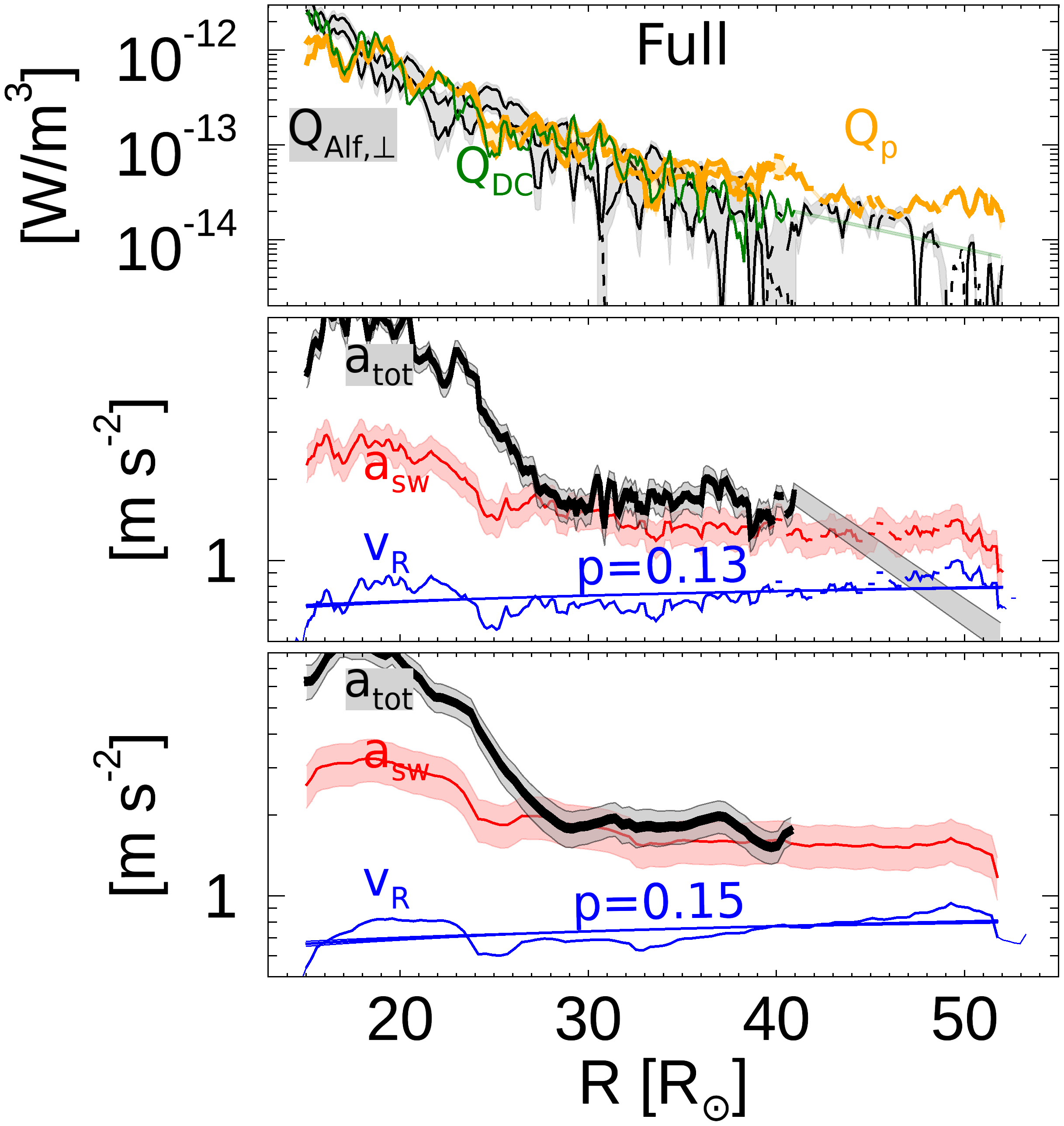}
\includegraphics[height=0.4\linewidth,trim={5cm .1cm 0cm 0.cm},clip=]{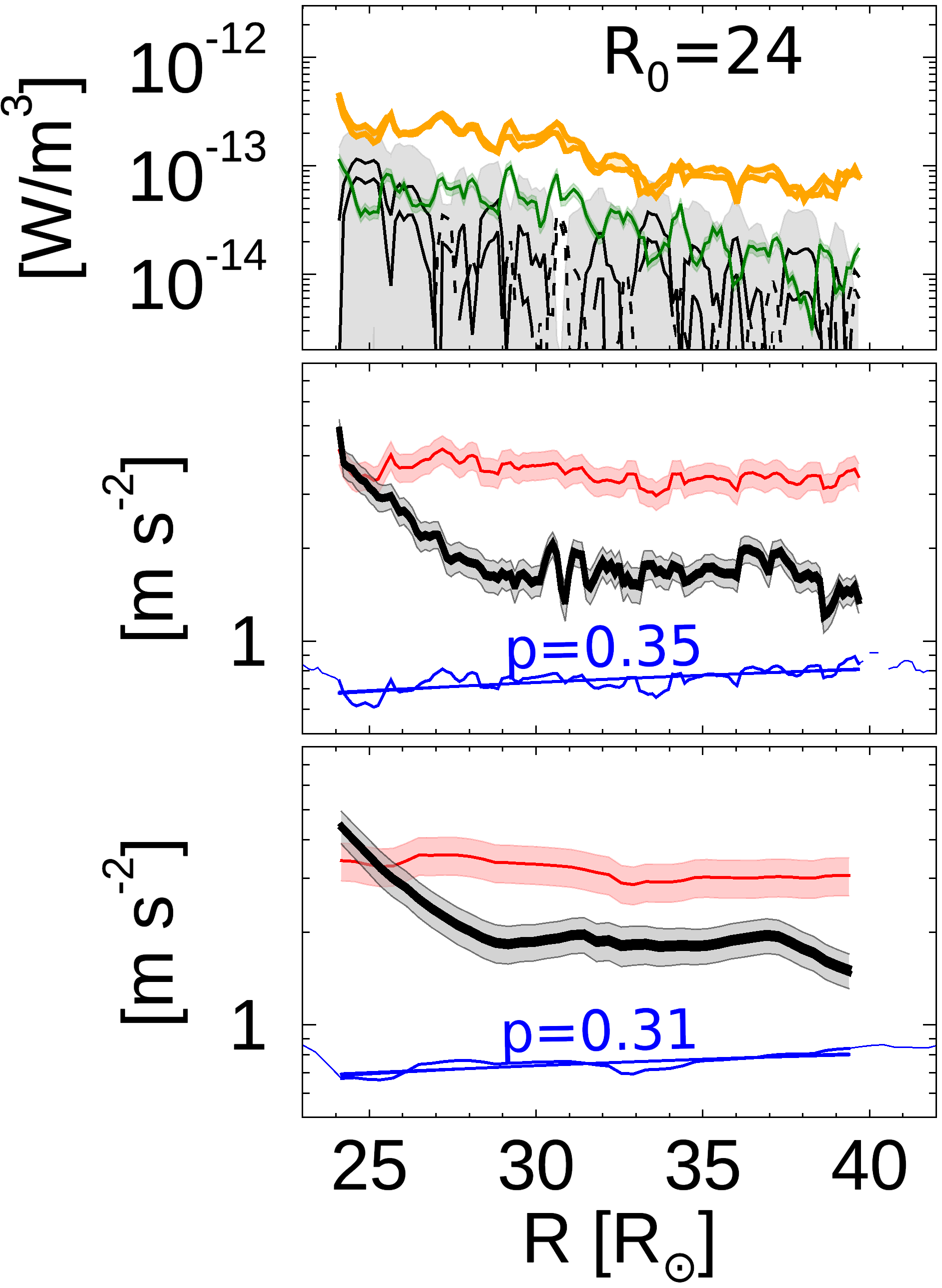}
\includegraphics[height=0.4\linewidth,trim={5cm .1cm 0cm 0.cm},clip=]{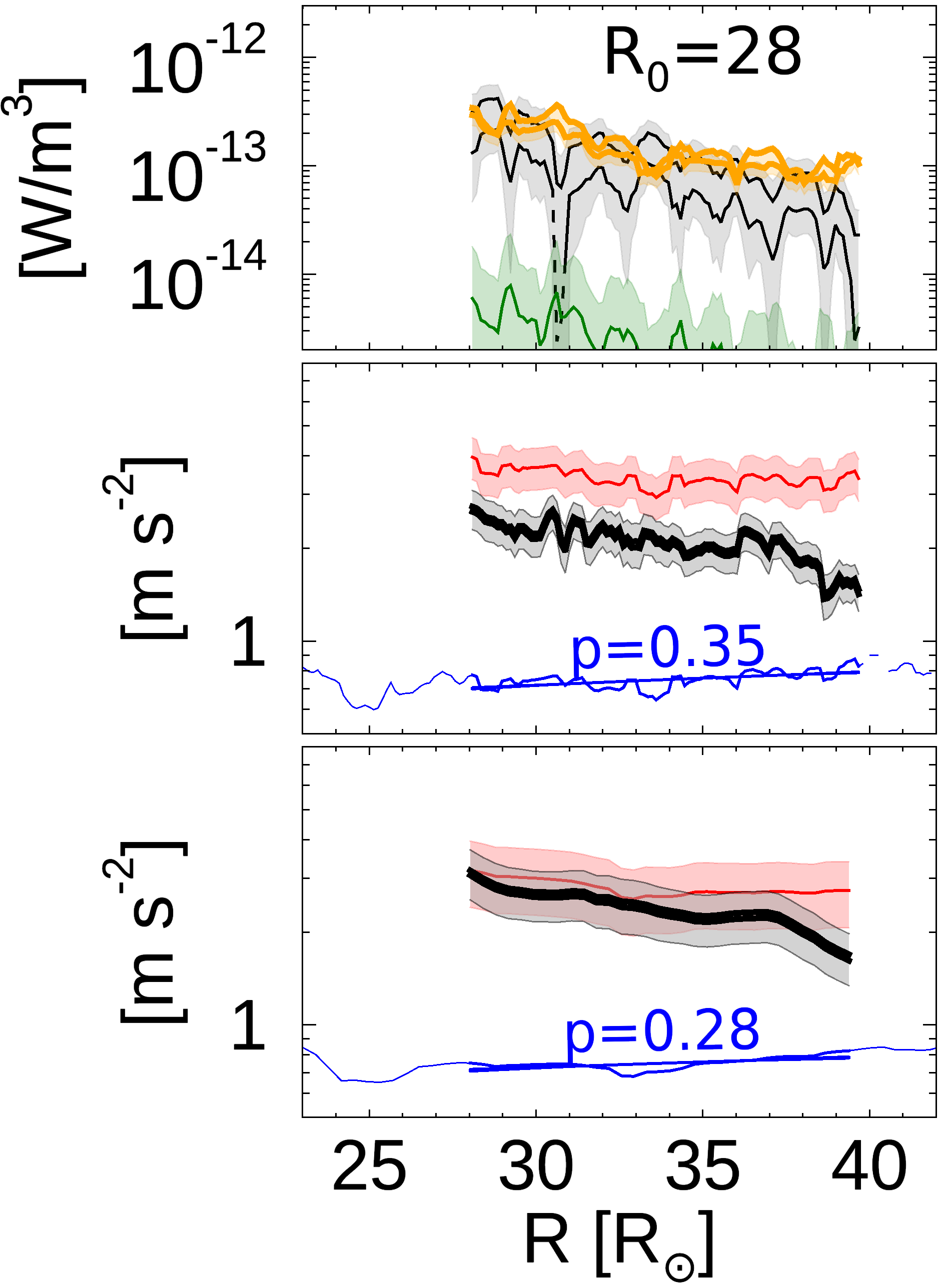}
\includegraphics[height=0.4\linewidth,trim={5cm .1cm 0cm 0.cm},clip=]{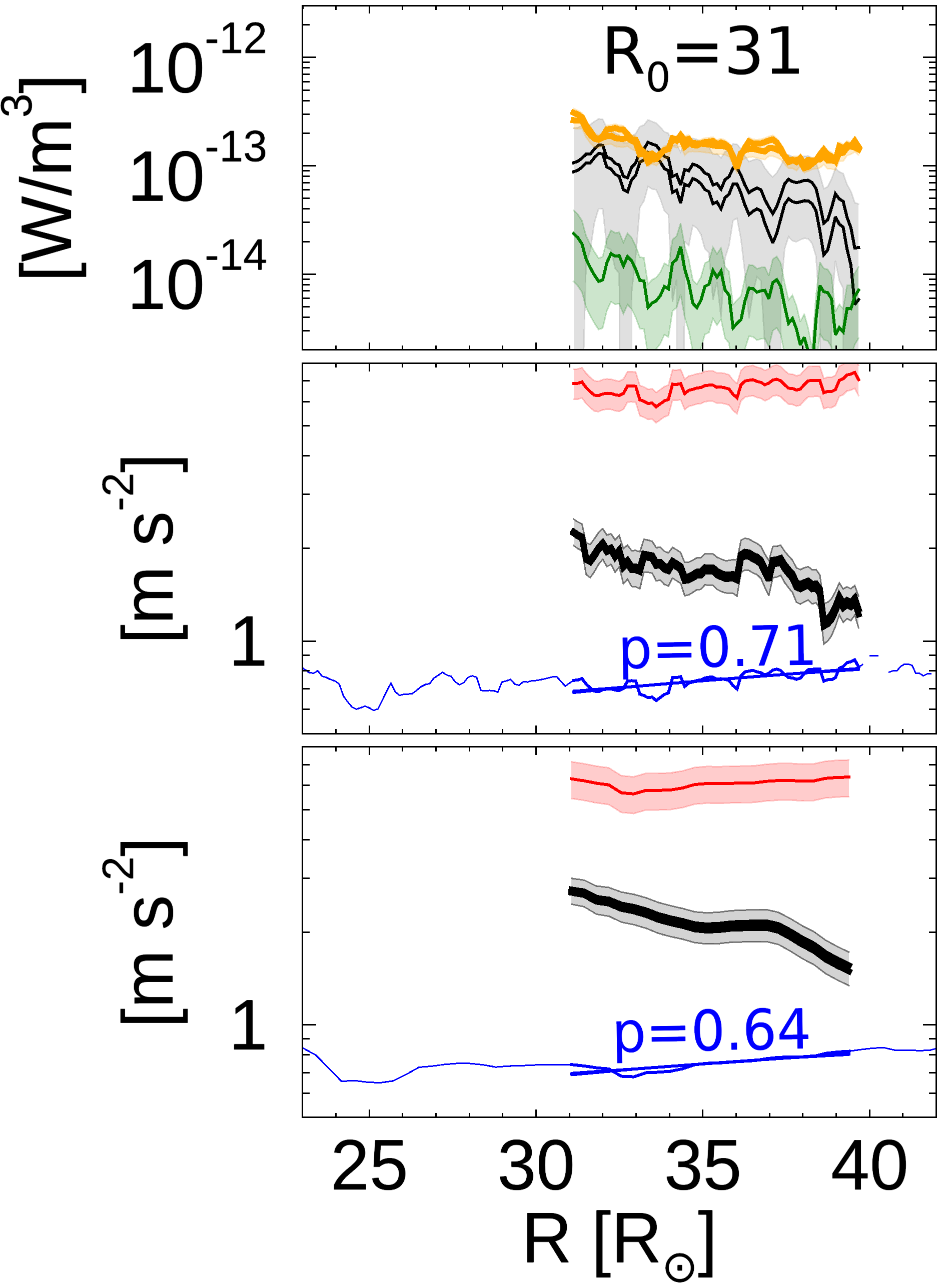}
\caption{Same as in figs.~\ref{fig:heat},~\ref{fig:acc} but considering the entire span of radial distances (left) and larger lower radial bounds ($R_0$).
Top Panel: comparison of heating estimate from energy conservation ($Q_{Alf,\bot}$, black) and turbulent phenomenology ($Q_{DC}$, green) with the expected proton and electron heating ($Q_e$, orange). 
Central and bottom panel: comparison of the solar wind acceleration ($a_{sw}$, red) with sum of thermal, fluctuations, torque, and gravity acceleration ($a_{tot}$, black) computed at scales $\tau=1$~hr (middle panels) and $\tau=6$~hr (bottom panels). Also plotted is the normalized velocity profile (thin blue line), the fitted range (thick blue line), and the power-law index $p$.}
\label{fig:range}
\end{figure*}

\section{Discussion}\label{sec:discuss}
In obtaining the heating and acceleration, we use the corotating part of the stream between 31 and 40$R_\odot$ and the analysis was performed at a scale $\tau=1$~hr. 
These choices are somewhat arbitrary, although the radial range was selected based on the geometry of the encounter and on the radial variation of conserved quantities (polar electric field, mass flux, angular momentum), and the scale identified as a trade-off between capturing large-scale fluctuations and sufficient statistics.
In fig.~\ref{fig:range} we show how the results based on the balance of energy and momentum change when the length of the analyzed interval and the scale for separating fluctuations and mean field are varied.
The radial range decreases from left to right, with the whole inbound orbit being represented in the leftmost panels and the previously selected interval in the rightmost panels.
The smallest distance in the interval is varied and indicated for reference as $R_0$, while the largest distance is set to $40~R_\odot$ (except in the first panel).
In the first row we compare the expected proton heating ($Q_e$), the heating resulting from the energy loss of fluctuations ($Q_{Alf,\bot}$), and the heating from turbulent phenomenology ($Q_{DC}$) that are plotted in orange, gray, and green, respectively.
Analyzing the inbound orbit as a whole, we recover the result of \citet{Bowen:2025aa}: $Q_e\approx Q_{DC}\approx Q_{Alf,\bot}$ at all distances, indicating that turbulent dissipation of fluctuations energy is responsible for the estimated proton heating.
However, such a large interval mixes streams detected at different longitudes in the frame of the Sun, questioning the validity of the above equality.
When shorter intervals are considered, $Q_p$ is not altered, while $Q_{DC}$ becomes smaller and insufficient. 
The value of $Q_{DC}$ is mainly controlled by the gradient of the Alfv\'en speed that is largely reduced when the closest streams ($R\lesssim24R_\odot$) are excluded from the analysis.
The heating from fluctuations is also largely insufficient, also taking negative values as indicated by the dashed black lines, when the interval includes entirely the density bump that we interpreted as a marker for shear interaction ($R_0=24R_\odot$).
As soon as only part of the density bump is included or when it is excluded ($R_0=28,~31R_\odot$), the heating from fluctuations returns consistent with the expected proton heating.

In the second row we plot the solar wind acceleration ($a_{sw}$) and the total acceleration ($a_{tot}$, sum of all contributions) with red and black colors, respectively. 
For the interval as a whole, the total acceleration paradoxically exceeds the solar wind acceleration at small distances, while they match in the range $R\in[28,40]R_\odot$. 
In this range the total acceleration is relatively constant, with values around 2-3$\mathrm{m~s^{-2}}$, whatever the length of the interval considered.
On the contrary the solar wind acceleration increases by a factor 7 when passing from left to right.
In other words, whenever $a_{sw}\approx a_{tot}$, it is because of the smaller solar wind acceleration and not because of the increased acceleration exerted by fluctuations.
Indeed, the power-law fit of the velocity profile that is used to evaluate $a_{sw}$ is largely influenced by the interval considered, as shown by the blue lines in each panel.

The velocity profiles highlight large-scale variations that are present also in temperature, density and magnetic field.
One may wonder how the above results change when such variations are considered as part of the fluctuations energy.
In the bottom panels we show the accelerations computed by separating mean values and fluctuations on a scale of 6~hr with 1hr sliding (we do not show the heating because no considerable difference is seen).
When considering the whole interval, there is basically no change in the results. 
For smaller radial ranges, the sum of the thermal acceleration and fluctuations acceleration (black) is not affected by the chosen scale.
On the contrary, the estimated solar wind acceleration (red) decreases because the smoothed velocity profiles (blue) have smaller power-law indexes compared to the 1hr averages in the central panels. 
Note, however, that the observed smoothing results also from the mixture of streams that are about 3$R_\odot$ apart.

\section{Summary and conclusions}\label{sec:conclu}
\begin{figure*}[ht!]
\sidecaption
  \includegraphics[width=12cm,trim={0.1cm .1cm 0cm 0.cm},clip=]{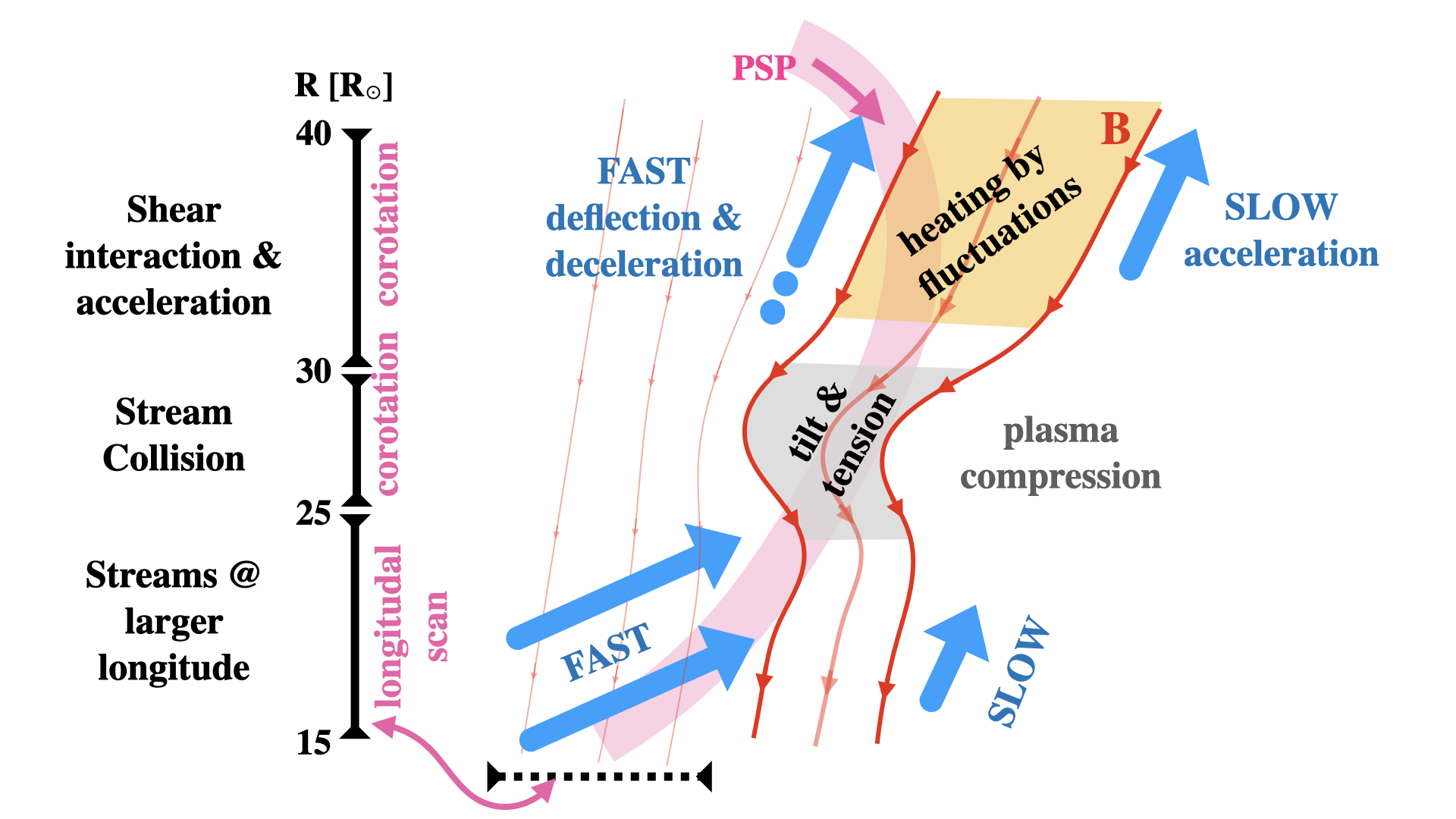}
\caption{Sketch of the flow geometry and interaction in the frame corortaing with the Sun during the inbound orbit of PSP (purple) in Encounter 10. 
Flow velocity vectors are in blue, magnetic field lines are in red, the flux tube detected by PSP is bounded by thicker lines. On the left we indicate the radial distances roughly corresponding to the quasi-longitudinal and the quasi-corotating scans by PSP in which the collision of streams and shear interaction occur.}
\label{fig:draw}
\end{figure*}
The analysis of PSP data during the inbound orbit in Encounter 10, when the spacecraft is spanning distances between 15 and 50$R_\odot$, returns a complex structure of the solar wind flow close to the Sun.
We identified two different stream structures, based on the velocity, density and magnetic field profiles computed on 1hr scale and on the kinematic back-mapping to 6$R_\odot$ in the frame rotating with Sun.
During PSP quasi longitudinal scan in the closest approach to the Sun, a collection of very fast streams with speed around 600km/s roughly originate from the same source and expand in a direction that strongly deviates from the radial one, with a tilt of about $-20^o$ (i.e. opposite to the Sun's rotation).
These streams point toward a slower stream that is detected during the corotation phase of PSP orbit between 30 and 40$R_\odot$, with speed around 450km/s and smaller tilt angle, about $-10^o$. 
Such geometry suggests that the two streams are colliding and their interface is roughly located between 25 and 30$R_\odot$ where density enhancement and large-scale magnetic distortion are seen.
Further out the slower corotating stream is heated and accelerates reaching the fast stream speed (see sketch in fig~\ref{fig:draw}).

Both stream have large amplitude magnetic fluctuations.
In the fast stream fluctuations are strongly Alfv\'enic with $\sigma_c\approx 0.9$ and field parallel magnetic and velocity components are negligible. 
In the corotating stream the cross helicity drops to 0.8, mainly because of the decorrelation of radial velocity and magnetic fluctuations, suggesting compressive fluctuations are generated by the stream interaction.
We computed the heat and the work on the bulk flow resulting from the loss of fluctuations energy with distance using a scale decomposition and the assumptions of stationary fields and incompressible fluctuations \citep{Hollweg:1974aa}.
We find that heating from fluctuations is barely sufficient to explain the non-adiabatic proton temperature profile measured in the corotating stream.
This result is consistent with \citet{Bowen:2025aa} who analyzed the same stream and the statistical analysis of \citet{Bourouaine:2024aa}.
In contrast, we find that turbulent phenomenology \citep{Chandran:2009tu} is a factor 10 too small in the range $R\in[31,40]R_\odot$. Agreement can be recovered only when longer intervals are considered, because the Alfv\'en speed and its radial variation is determined by a mixture of different streams.
We also find that work and heat are almost equally partitioned, similarly to the class of winds with proton energies $\in[1500-2000]~\mathrm{eV}$ (the second fastest winds) in the statistical analysis of \citet{Halekas:2023uc}.

We finally evaluated both the direct acceleration caused by work of fluctuations (via gradient of the pressure of fluctuations) and the indirect acceleration by heat deposition (via gradient of the thermal pressure). 
Surprisingly, their sum is insufficient to explain the velocity increase observed in the corotating stream, at variance with conclusions drawn from the energy balance \citep{Halekas:2023uc,Rivera:2024aa}.
This result is confirmed by further analysis on different range of distances and at larger timescale (6~hr), although under the limiting assumptions of incompressible fluctuations and stationary fields.
Indeed, the increase of radial velocity fluctuations in the corotating stream indicates compressibility may play a role in its acceleration.
Also, the presence of switchback patches in PSP data during corotation in this same Encounter, led \citet{Shi:2020aa} to suggest they originate from temporal variation at the source rather than from spatial variation on supergranulation scale \citep{Bale:2021aa,Fargette:2021aa}.
Preliminary simulations of one-dimensional solar wind with time dependent input of fluctuations indicate that within the observational constrains given by PSP data, the compressibility does not affect our main conclusion on insufficient radial acceleration. Time dependence of mean field is more difficult to evaluate, indeed we observe transient strong acceleration when the input spectrum of fluctuations contains sufficiently small frequencies.

In conclusion, our work suggests that the acceleration of the corotating stream between 31 and 40~$R_\odot$ cannot be explained with energy and momentum exchange between the bulk flow and fluctuations along the radial direction, but (at least) two-dimensional dynamics must come into play.
The small density enhancement and large-scale magnetic field distortion that we observe at the interface between the two streams in PSP orbit are characteristic signatures of an ongoing shear interaction \citep{Grappin:1996wd,Landi:2006wo,Shi:2020aa}. 
An order of magnitude estimate of the distances at which shear interaction is at work yields $\Delta R_{sh}=\Delta R_\bot V_{cor}/\Delta V\approx 5-10R_\odot$, where $V_{cor}\approx 500~\mathrm{km~s^{-1}}$ is the velocity of corotating stream,
$\Delta V\approx 100~\mathrm{km~s^{-1}}$ is the velocity shear and its width is $\Delta R_\bot\approx1-2 R_\odot$ from the kinematic trajectories in the frame rotating with the Sun (the distance between the blue and the gray trajectories labelled 3 and 4 in fig.~\ref{fig:orbit}d).
This range is comparable to that of the corotating stream ($\approx9~R_\odot$).
Interestingly, \citet{Horbury:2023aa} showed that switchback patches are strongly correlated with microstreams at $20~R_\odot$, both having a characteristic angular scale of $5^o$ in their power spectrum, but their signature disappeared at $200~R_\odot$. Switchback patches persisted only in the spectrum of magnetic deflections (with respect to the mean field) and a slightly larger characteristic scale appeared in the density spectrum.
Although further observations and dedicated numerical experiments are needed, we suggest that the uniformity of stream observed already at 0.3 au in Helios data is achieved via shear interaction at heliocentric distances of about $40~R_\odot$ or below. Shear interaction is efficient because streams are non-radial and it may leave its imprint at larger distances in density enhancement and large-scale magnetic field deflections.

\begin{acknowledgements} 
This research was partially funded by the European Union - Next Generation EU - National Recovery and Resilience Plan (NRRP) - M4C2 Investment 1.4 - Research Programme  CN00000013 ``National Centre for HPC, Big Data and Quantum Computing'' - CUP B83C22002830001 and by the European Union - Next Generation EU - National Recovery and Resilience Plan (NRRP)- M4C2 Investment 1.1- PRIN 2022 (D.D. 104 del 2/2/2022) - Project `` Modeling Interplanetary Coronal Mass Ejections'', MUR code 31. 2022M5TKR2,  CUP B53D23004860006.
Views and opinions expressed are however those of the author(s) only and do not necessarily reflect those of the European Union or the European Commission. Neither the European Union nor the European Commission can be held responsible for them.
This research was supported by the International Space Science Institute (ISSI) in Bern, through ISSI International Team project n.560 and n.591.
L.F. is supported by the Royal Society University Research Fellowship No. URF/R1/231710.
This work has been supported by Agenzia Spaziale Italiana (ASI) in the framework of the project ``Radial Evolution of large- and KInetic-scale Processes in the Expanding Solar wind (REKIPES)'' C83C25000900005.
The authors acknowledge useful discussions with Lorenzo Matteini and Roland Grappin.
\end{acknowledgements}

\bibliographystyle{aa}
\bibliography{bibcor}

\onecolumn

\appendix
\section{Alfv\'enic and perpendicular model}\label{sec:app1}
Our derivation follows closely \citet{Hollweg:1974aa}, the only difference is that equations are reduced to the Alfv\'enic limit (the original Hollweg work) and to the perpendicular limit only at the very last stage.

We start by assuming stationary fields so that the MHD equations in conservation form can be written as,
\ba
\divv\left(\rho \vect{U}\right)&=&0 \label{eq:cont}\\
\divv\left[\rho\vect{UU}+
  \left(P+\frac{B^2}{4\pi\rho}\right)\vect{I}
  -\frac{\vect{BB}}{4\pi}\right]&=&\rho\vect{g} \label{eq:mom}\\
\divv\left[\left(\frac{1}{2}\rho U^2+\frac{5}{2}P\right)\vect{U}+\vect{q}+\vect{S}\right]&=&\rho\vect{U}\cdot\vect{g} \label{eq:ene}
\ea
where, as customary, $\rho,~\vect{U},~\vect{B}$ stand for density, velocity filed, magnetic fields, $\vect{q}$ is the thermal conduction, $\vect{S}=(c/4\pi)\vect{E}\times\vect{B}$ is the Poynting flux, $\vect{g}$ is the gravitational force and we have assumed isotropic pressure $\vect{P}=P\vect{I}$ with $\vect{I}$ being the unit tensor.

We separate each variable in a mean and fluctuating part, the latter having vanishing mean.
Fluctuations are further assumed to be incompressible, resulting in vanishing density and pressure fluctuations, so that finally the decomposition reads,
\ba
\vect{U}&=&\vect{U_0}+\vect{u}\;\;\mbox{with}\;\;\vect{U_0}=\lla\vect{U}\rra\;\mbox{and}\;\lla\vect{u}\rra=0 \\
\vect{B}&=&\vect{B_0}+\vect{b}\;\;\mbox{with}\;\;\vect{B_0}=\lla\vect{B}\rra\;\mbox{and}\;\lla\vect{b}\rra=0 \\
       P&=&P_0=\lla P\rra \\
       \rho&=&\rho_0=\lla \rho\rra 
\ea
Inserting these definitions in eqs.~\eqref{eq:cont}-\eqref{eq:ene} and averaging one obtains the following set of equations:
\ba
&\divv&\left[\rho_0\vect{U_0}\right]=0\label{eq:mass}\\
&\divv&\left[\rho_0\vect{U_0U_0}+\rho_0\lla\vect{u}\vect{u}\rra 
  + \left(P_0+\frac{B_0^2}{8\pi}\right)\vect{I}
    -\frac{\vect{B_0B_0}}{4\pi}
    -\frac{\lla\vect{bb}\rra}{4\pi}\right]-\rho_0\vect{g}=
  -\grad\frac{\lla b^2\rra}{8\pi}\label{eq:mom2}\\
&\divv&\left\{\frac{1}{2}\rho_0\left[\left(U_0^2+\lla u^2\rra\right)\vect{U_0}
       +\lla u^2\vect{u}\rra+2\lla(\vect{U_0}\cdot\vect{u})\vect{u}\rra\right]
    +\frac{5}{2}P_0\vect{U_0}\right\}+\label{eq:ene2}\\
&\divv&\frac{1}{4\pi}\bigg[\left(B_0^2+\lla b^2\rra\right)\vect{U_0}
 -\left(\vect{U_0}\cdot\vect{B_0}+\lla\vect{u}\cdot\vect{b}\rra\right)\vect{B_0}
\lla(b^2)\vect{u}+2(\vect{B_0}\cdot\vect{b})\vect{u}
  -(\vect{u}\cdot\vect{b})\vect{b} -(\vect{u}\cdot\vect{B_0})\vect{b}
  -(\vect{U_0}\cdot\vect{b})\vect{b}\rra\bigg]\nonumber\\
&=&\rho_0\vect{U_0}\cdot\vect{g}\nonumber
\ea 
where the second term on the l.h.s of eq.~\eqref{eq:ene2} is the decomposition of the Poynting flux.
For completeness, requiring stationary magnetic field in the induction equation one has
\be
\curl\left(\vect{U_0}\times\vect{B_0}\right)=-\curl\lla \vect{u}\times\vect{b}\rra.\label{eq:ind}
\ee
Taking the scalar product of eq.~\eqref{eq:mom2} with $\vect{U_0}$ to eliminate gravity from eq.~\eqref{eq:ene2} and using eq.~\eqref{eq:ind},
one finally obtains an equation for the energy balance of fluctuations (last equality below),
\be
\divv\left[\left(\frac{3}{2}P_0\right)\vect{U_0}+\vect{q}\right]+P_0(\divv\vect{U_0})=Q=-\divv \phi_f -W_f,
\label{eqa:enbal}
\ee
where $Q>0$ is the volumetric heating that results from damping ($-Q<0$) of fluctuations energy and increases the thermal energy of the solar wind, the flux of fluctuations energy is 
\be
\phi_f=\left[\frac{1}{2}\rho_0\lla u^2\rra+
             \lla\frac{b^2}{4\pi}\rra\right]\vect{U_0}
       -\left[\frac{\lla \vect{u}\cdot\vect{b}\rra}{4\pi}\right]\vect{B_0}+
\frac{1}{2}\rho_0\bigg[\lla 2(\vect{u}\cdot\vect{U_0})\vect{u}\rra +
     \lla(u^2)\vect{u}\rra\bigg]
-\frac{1}{4\pi}\bigg[\lla (\vect{b}\cdot\vect{U_0})\vect{b}\rra +
     \lla(\vect{u}\cdot\vect{B_0})\vect{b}\rra
     +\lla(\vect{u}\cdot\vect{b})\vect{b}\rra\bigg],
\label{eqa:flux}
\ee
and the work done by fluctuations on the solar wind is
\be
W_f=-\vect{U_0}\cdot\left[\divv\lla\rho_0\vect{uu}
                             -\frac{\vect{bb}}{4\pi}\rra\right]
-\vect{B_0}\cdot\left[\frac{\curl\lla\vect{b}\times\vect{u}\rra}{4\pi}\right]
-\vect{U_0}\cdot\grad\lla\frac{b^2}{8\pi}\rra.
\label{eqa:work}
\ee

We now assume that in the frame corotating with the Sun, $\vect{U_0}$ and $\vect{B_0}$ are aligned, and derivative can be taken with respect to the curvilinear coordinate (indicated with $R$ below).

In the Alfv\'en model we assume fluctuations are perfectly correlated and aligned, 
$\vect{u}=\vect{b}/\sqrt{4\pi\rho_0}$ for $B_R<0$ pointing toward the sun ($\vect{u}=-\vect{b}/\sqrt{4\pi\rho_0}$ for $B_R>0$).
We also assume that the magnetic field magnitude is not changed by fluctuations, $\delta|B|=const$, which implies
$\drle\lla \vect{b}\cdot\vect{B_0}\rra=- \drle\lla b^2/2\rra$.
In eq.~\eqref{eqa:flux} the last two terms vanish, while in eq.~\eqref{eqa:work} the first and second term vanish.
Expressing the divergence in spherical coordinates one finally obtains from eq.~\eqref{eqa:enbal} the fluctuations energy balance,
\be
\frac{1}{A}\dre\left\{A\left[U_0
\frac{3}{2}\frac{\lla b^2\rra}{4\pi}+
    V_{A0}\frac{\lla b^2\rra}{4\pi} \right]\right\}
+\left[-U_0\dr{}{\lla \frac{b^2}{8\pi}\rra}\right] 
= -Q
\;\;\mbox{\textit{(Alfv\'en model)}},
\label{eqa:alfMass}
\ee 
where the terms in square brackets are $F_{Alf}$ and $W_{Alf}$, eqs.~\eqref{eq:a_alf}-\eqref{eq:f_alf}.
In the Perpendicular model we assume fluctuations to be perpendicular to the mean velocity and magnetic field. 
Again the last two terms in eq.~\eqref{eqa:flux} vanish, while in eq.~\eqref{eqa:work} only the second term vanishes.
Using the divergence of the tensor in spherical coordinates one finally obtains 
\be
\frac{1}{A}\dre\left\{A\left[\rho_0U_0
\frac{\lla u_\bot^2\rra}{2}+U_0\frac{\lla b_\bot^2\rra}{4\pi}
   -B_0\frac{\vect{u_\bot}\cdot\vect{b_\bot}}{4\pi}\right]\right\}+
\left[-U_0\dr{}{\lla \frac{b_\bot^2}{8\pi}\rra}
        +U_0\frac{A'}{2A} \left(\rho_0\lla u_\bot^2\rra 
        -\frac{\lla b_\bot^2\rra}{4\pi}\right)\right]
= -Q
\;\;\mbox{\textit{(Perpendicular model)}},
\label{eqa:perpMass}
\ee
where now the terms in square brackets are $F_\bot$ and $W_\bot$, eqs.~\eqref{eq:a_bot}-\eqref{eq:f_bot}.

Note that in both equations the mass flux can be factorized to yield equivalent forms,
\be
\dre\left[
\frac{3}{2}\frac{\lla b^2\rra}{4\pi\rho_0}+
    \frac{V_{A0}}{U_0}\frac{\lla b^2\rra}{4\pi\rho_0} \right]
+ \left[-\frac{1}{8\pi\rho_0}\dr{\lla b^2\rra}\right] 
= -\frac{Q}{\rho_0U_0}
\;\;\mbox{\textit{(Alfv\'en model)}},
\label{eqa:alf}
\ee 
and
\be
\dre\left[
\frac{\lla u_\bot^2\rra}{2}+\frac{\lla b_\bot^2\rra}{4\pi\rho_0}
   -\frac{B_0}{U_0}\frac{\vect{u_\bot}\cdot\vect{b_\bot}}{4\pi\rho_0}\right]
+ \left[-\frac{1}{8\pi\rho_0}\dr{\lla b_\bot^2\rra}
        +\frac{A'}{2A} \left(\lla u_\bot^2\rra 
        -\frac{\lla b_\bot^2\rra}{4\pi\rho_0}\right)\right]
= -\frac{Q}{\rho_0U_0}
\;\;\mbox{\textit{(Perpendicular model)}},
\label{eqa:perp}
\ee
but we prefer to use eqs.~\eqref{eqa:alfMass}-\eqref{eqa:perpMass} because the mass flux is not exactly constant.

We conclude by showing that for an axisymmetric solar wind, both the Alfv\'en and the Perpendicular models lead to the same invariants originally obtained by \citet{Weber:1967aa}.
Let us assume that the stationary mean fields depend only on the radial coordinate $r$ and have vanishing $\theta$ component, that is, $\vect{U_0}=V_R\hat{\vect{r}}+V_T\hat{\vect{\phi}}$ and $\vect{B_0}=B_R\hat{\vect{r}}+B_T\hat{\vect{\phi}}$.
Eq.~\eqref{eq:mass} is obtained under the assumption of incompressible fluctuations that is common to both fluctuations models, leading to the conservation of the mass flux 
\be
\dot{M}=\rho_0V_RA=const. \label{eq:massw}
\ee
In the l.h.s. of the $\phi$ component of the momentum equation, eq.~\eqref{eq:mom2}, the terms involving fluctuations vanish for the  Alfv\'enic model ($\vect{u}//\vect{b}$) and for the Perpendicular model ($u_R=B_R=0$). Using the divergence of a tensor for a flux tube with cross section $A(r)$, the conservation of angular momentum reads
\be
\mathcal{L}=\sqrt{A}(V_T-\frac{B_R}{4\pi\rho V_R}B_T)=const.\label{eq:momw}
\ee
Similarly, in the $\phi$ component of the stationary induction equation for the mean magnetic field,  eq.~\eqref{eq:ind}, the r.h.s. vanishes for the Alfv\'enic and the Perpendicular model.
Thus, accounting for the non-radial expansion of the flux tube, the derivative along $r$ of the polar component of the electric field multiplied by $\sqrt{A}$ must vanish, yielding
\be
E^\theta=\sqrt{A}(V_RB_T-V_TB_R)=const.\label{eq:Maxwellw}
\ee

\end{document}